\documentclass{emulateapj}

\newcommand{\kms}{km~s$^{-1}$}
\newcommand{\helium}{\ion{He}{1}~$\lambda$10830}
\newcommand{\heopt}{\ion{He}{1}~$\lambda$5876}
\newcommand{\pgamma}{P$\gamma$}

\newcommand{\msunyr}{$M_\sun$~yr$^{-1}$}
\newcommand{\nod}{\nodata}

\newcommand{\heupper}{$2p~{}^3P^o$}
\newcommand{\helower}{$2s~{}^3S$}
\newcommand{\vred}{$V_{\rm red}$}
\newcommand{\vesc}{$V_{\rm esc}$}

\slugcomment{Accepted by ApJ, to appear 2008 November 10}

\begin{document}

\title{Redshifted Absorption at He I $\lambda$10830 as a Probe of the \\Accretion Geometry of T Tauri Stars}
\author{William Fischer\altaffilmark{1,4}, John Kwan\altaffilmark{1}, Suzan Edwards\altaffilmark{2,4}, Lynne Hillenbrand\altaffilmark{3,4}}
\altaffiltext{1} {Five College Astronomy, University of
 Massachusetts, Amherst, MA 01003,
 wfischer@astro.umass.edu, kwan@astro.umass.edu}
\altaffiltext{2}{Five College Astronomy, Smith College,
 Northampton, MA 01063, sedwards@smith.edu}
\altaffiltext{3} {Dept.\ of Astronomy, California
Institute of Technology, Pasadena, CA 91125,
lah@astro.caltech.edu}
\altaffiltext{4}{Visiting Astronomer, Keck Observatory}
 
\begin{abstract}
We probe the geometry of magnetospheric accretion in classical T Tauri stars by modeling red absorption at \helium\ via scattering of the stellar and veiling continua.  Under the assumptions that the accretion flow is an azimuthally symmetric dipole and helium is sufficiently optically thick that all incident 1-$\micron$ radiation is scattered, we illustrate the sensitivity of \helium\ red absorption to both the size of the magnetosphere and the filling factor of the hot accretion shock. We compare model profiles to those observed in 21 CTTS with subcontinuum redshifted absorption at \helium\ and find that about half of the stars have red absorptions and 1-$\micron$ veilings that are consistent with dipole flows of moderate width with accretion shock filling factors matching the size of the magnetospheric footpoints. However, the remaining 50\% of the profiles, with a combination of broad, deep absorption and low 1-$\micron$ veiling, require very wide flows where magnetic footpoints are distributed over 10-20\% of the stellar surface but accretion shock filling factors are $<1\%$. We model these profiles by invoking large magnetospheres dilutely filled with accreting gas, leaving the disk over a range of radii in many narrow ``streamlets'' that fill only a small fraction of the entire infall region. In some cases accreting streamlets need to originate in the disk between several $R_*$ and at least the corotation radius. A few stars have such deep absorption at velocities $>0.5~V_{\rm esc}$ that flows near the star with less curvature than a dipole trajectory seem to be required.
\end{abstract}


\section{INTRODUCTION}

The classical T Tauri stars (CTTS) are optically revealed, low-mass, pre--main-sequence stars that accrete material from a circumstellar disk and have a well-defined connection between accretion and outflow \citep[HEG hereafter]{har95}.  Accretion from the disk to the star is thought to be guided by the stellar magnetosphere, where  a sufficiently strong magnetic field truncates the disk at several stellar radii and material follows  field lines that direct it to the stellar surface at high latitudes \citep{gho78,kon91,col93,shu94}.  Magnetospheric accretion controls the star-disk interaction, and an improved understanding of this process will shed light on outstanding issues in the innermost 10~$R_*$ of CTTS systems, such as the regulation of stellar angular momentum and the launching of the inner wind.

Leading diagnostics of CTTS accretion include the optical/UV continuum excess and the profiles of permitted emission lines \citep{bou07a}.  Kinematic evidence for infalling gas in CTTS began with the discovery that some CTTS show inverse P Cygni structure in upper Balmer lines extending to velocities of several hundred \kms\ \citep{wal72}.  Later, more sensitive surveys found that redshifted absorption components are relatively common in some lines, especially in the upper Balmer and Paschen series \citep{edw94,ale00,fol01}.  Although redshifted absorption extending to several hundred \kms\ clarifies that material accretes in free-fall from at least several stellar radii, it has been the success of radiative transfer modeling of line formation in magnetospheric accretion flows in a key series of papers culminating with \citet{muz01} that has provided the strongest underpinning for this phenomenon.  Under the assumption of an aligned, axisymmetric dipole, the models have had reasonably good success in reproducing the general morphology of hydrogen profiles and emission fluxes in some stars.  The complementary assessment of accretion rates follows from interpreting the SED of the optical/UV excess, which has been successfully modeled for wavelengths shortward of 0.5~$\micron$ as arising in a hot accretion shock, where accreting material impacts the stellar surface after free-fall along funnel flows coupled to the disk \citep[CG hereafter]{joh00,cal98}. To match the observations, the accretion shock filling factor is less than 1\% in most cases but can climb to 10\% in a few of the most active accretors.  The derived accretion rates range from $10^{-10}$ to $10^{-6}$ \msunyr\ with a median of $10^{-8}$~\msunyr\ \citep{cal00}.

Additionally, Zeeman broadening of unpolarized CTTS photospheric lines indicates mean surface field strengths in the range 1-3~kG \citep{joh07}, sufficiently strong to induce disk truncation and drive funnel flows.  However, these strong surface fields are not predominantly dipolar, as photospheric lines show  only weak net circular polarization implying dipole components an order of magnitude smaller \citep{val04,yan07}. Nevertheless, an extended dipole component is inferred for the accretion flow, since the same authors find significant circular polarization in the narrow component of the \heopt\ emission line, thought to be formed in the accretion shock at the base of the funnel flow \citep{ber01}.

The evidence for magnetospheric accretion is thus compelling; however, the topology of the magnetosphere, the geometry of the accretion flow, and the disk truncation radius remain topics of considerable investigation, since their configuration impacts processes for angular momentum regulation and wind launching.  One form of angular momentum regulation, known as disk locking, invokes a spin-up torque from accreting material just inside the corotation radius balanced by a spin-down torque at larger radii \citep{col93}. This approach has been questioned by \citet{mat05,mat08a,mat08b}, who instead suggest accretion-powered stellar winds as a more likely means for stellar spindown, which must occur simultaneously with magnetospheric accretion from the disk (see also \citealt{sau94}).  Alternatively the X-wind model \citep{shu94} originally featured a narrow annulus of star-disk coupling close to the corotation radius, where closed field lines develop funnel flows and open field lines drive a centrifugal outflow that carries away angular momentum from accreting material, thus inhibiting stellar spin-up.  The flexibility of this model to maintain its basic properties in the face of complex magnetospheric accretion geometries has recently been demonstrated by \citet{moh08}. If the stellar and disk fields are parallel, then an intermittent outflow can develop via a Reconnection X-wind, which removes angular momentum from the star as well as the inner disk \citep{fer06}.

Evidence for non-aligned fields coupled with complex accretion geometries is mounting, coming from a variety of recent studies.  Using time-resolved spectropolarimetry of the mildly accreting CTTS V2129 Oph, \citet{don07} used Zeeman detections from both photospheric features and emission lines from the accretion shock to construct a Doppler tomographic map of the magnetic topology on the stellar surface.  The dominant field on the star is a misaligned octupole, and accretion is confined largely to a high-latitude spot, covering $\le5$\% of the stellar surface. These authors also attempt to reconstruct the 3D field geometry out to the disk interaction region and suggest that the large-scale field funneling accreting material from the disk is more complex than a simple dipole. However, such extrapolation techniques, while tantalizing, must be applied with caution at present, since there are numerous uncertainties in the reconstruction process \citep{moh08}. Numerical magnetohydrodynamic (MHD) simulations of star-disk interactions demonstrate that non-axisymmetric funnel flows arise if the stellar field is tipped by only a few degrees relative to the rotation axis, breaking into two discrete streams under stable accretion conditions \citep{rom03}. If the accretion rate is sufficiently high, accretion is predicted to proceed through equatorial ``tongues'' that can push apart field lines \citep{rom08}. Similarly, accretion spots can appear at a wide range of latitudes, including equatorial belts.  Such behavior has been observed in models featuring accretion along quadrupolar as well as dipolar field lines \citep{lon07,lon08}, accretion along field lines extrapolated from surface magnetograms \citep{gre06}, and accretion mediated by a dynamo-generated disk magnetic field \citep{von06}.

Observational signatures for misaligned dipoles are being explored in radiative transfer models for hydrogen line formation in funnel flows, with subsequent predictions for rotationally modulated profile variations. Initially \citet{sym05} presented radiative transfer models of hydrogen lines featuring curtains of accretion covering a limited extent in azimuth in geometries consistent with aligned dipoles.  Their model profiles exhibit certain characteristics of the observed line profile variability, such as rotationally modulated line strengths and the appearance of red absorption components at certain phases and inclinations, but the predicted level of variability is higher than observed.  More recently, \citet{kur08} applied a radiative transfer code for H line formation to the 3-D output from the MHD simulations of \citet{rom03,rom04}, which prescribe the geometry, density, and velocity of two-armed accretion streams that result from dipoles misaligned with the rotation axis by angles ranging from 10$^\circ$ to 90$^\circ$.  Applying temperatures similar to those from the \citet{muz01} axisymmetric models, they were able to reproduce some of the trends in continuum and profile variability from models. One of the larger discrepancies in comparing the model profiles to observed ones is that the model profiles for Paschen and Brackett lines are a factor of two narrower than the mean value observed by \citet{fol01}. The problem is likely more complex than simply finding another line broadening mechanism, since Paschen~$\gamma$ has recently been shown to have line widths that are correlated with the 1-$\micron$ continuum excess, in the sense that the narrowest lines are found among objects with the lowest disk accretion rates \citep[EFHK hereafter]{EFHK}. Evidence that some of the hydrogen emission is formed in the accretion shock rather than the funnel flow is now clearly demonstrated from the discovery of circular polarization in the core of Balmer lines \citep{don08}. The implication is that hydrogren lines are not necessarily a definitive means for probing the properties of funnel flows, so additional probes are desirable.

In this paper, we explore a different means of diagnosing the geometry of the accretion flow, making use of the redshifted subcontinuum absorption in the \heupper\ $\rightarrow$ \helower\ transition of neutral helium ($\lambda10830$), recently demonstrated to be a very sensitive probe of both outflowing gas in the inner wind and infalling gas in the funnel flow due to its frequent display of blue and red absorptions (EFHK).  A subcontinuum absorption feature is a more tell-tale diagnostic of a kinematic flow than an emission profile, since its position (blue or red) indicates the direction of the flow, its width indicates the range of line-of-sight velocities in the flow, and its depth at a particular velocity indicates the fraction of the continuum (stellar plus veiling) occulted by material moving at that velocity.  In the case of a red absorption, the absorption depth signals the fraction of the stellar surface covered by the funnel flow at each velocity, making it an effective probe of the CTTS accretion geometry.

This is the third in a series of papers about 1-$\micron$ diagnostics of accretion and outflow in CTTS.  The first, EFHK, presented 1-$\micron$ spectra from 38 CTTS including \helium\ profiles, \pgamma\ profiles, and measurements of the continuum excess ``veiling''  in the 1-$\micron$ region.  The second paper, \citet[KEF hereafter]{kwa07}, modeled blueshifted absorption components at \helium, which appear in about three quarters of the sample, and found that while some stars have winds best explained as arising from the inner disk, others require an outflow moving radially away from the star in an accretion-powered ``stellar'' wind.  In this paper, we analyze redshifted subcontinuum absorption at \helium\ in 21 CTTS that present red absorption in at least one observation.  The following section describes the sample selection, data acquisition, and data reduction.  Section 3 presents the data and discusses variability.  In Section 4, we present model scattering profiles that arise in a dipolar flow geometry, show that they explain only a fraction of the red absorptions, and explore modifications to a dipolar flow that better explain the remaining observations.  Discussion and conclusions follow in Sections 5 and 6.

\section{SAMPLE AND DATA REDUCTION}

In this paper we focus on the 21 of 38 CTTS included in EFHK that display redshifted subcontinuum absorption at \helium\ at least once in a multi-epoch observing program with Keck NIRSPEC.  It includes spectra presented in EFHK acquired in November 2001 and November 2002, when 8 of the 38 CTTS were observed twice and 1 on three occasions.  It also includes 33 additional spectra of 24 objects from that study, taken in 2005, 2006, and 2007.  In EFHK, 19 out of 38 CTTS showed red absorption at \helium.  In the subsequent observing runs, \helium\ red absorption was seen in 2 additional stars.  Thus, the 21 of 38 CTTS (55\%) that have shown subcontinuum red helium absorption in at least one spectrum of 81 acquired between 2001 and 2007 form the sample for this paper.  Among the 21 stars with helium red absorption, 12 were observed more than once, with 6 observed twice, 1 observed three times, 4 observed four times, and 1 observed six times. The EFHK sample was assembled to span the full range of mass accretion rates observed for CTTS, from less than 10$^{-9}$ to $\sim 10^{-6}$~\msunyr, with a median rate of 10$^{-8}$~\msunyr. Most of them are from the Taurus-Auriga star-forming region and have spectral types of K7 to M0. The subset of 21 stars that are the focus of this paper are identified in Table~\ref{t.sample}, along with their spectral types, masses, radii, rotation periods, median veilings $r_V$ at 0.57 $\micron$, and mass accretion rates from the literature.  We have included $r_V$ only for the 29 sources in common with HEG, obtained more than a decade earlier.  We also list the number of observations for each star.  

\begin{deluxetable*}{lcccccccccc}
\tablecaption{21 of 38 CTTS with Subcontinuum Red Absorption at \helium\label{t.sample}}
\tablewidth{6in}
\tablehead{\colhead{Object} & \colhead{Sp Type} & \colhead{$M_*$} & \colhead{$R_*$} & 
\colhead{\vesc} & \colhead{$P_{\rm rot}$} & \colhead{$R_{\rm co}$} & \colhead{$\left<r_V\right>$} & \colhead{$\log \dot{M}_{\rm{acc}}$} & \colhead{Ref} & \colhead{$N_{\rm obs}$} \\ \colhead{(1)} & \colhead{(2)} & \colhead{(3)} & \colhead{(4)} & \colhead{(5)} & \colhead{(6)} & \colhead{(7)} & \colhead{(8)} & \colhead{(9)} & \colhead{(10)} & \colhead{(11)}}
\startdata
AA Tau\dotfill   & K7	& 0.70 & 1.75 & 390  & 8.22 & 8.7  & 0.32 & -8.5    & 12,9,5,9     & 4 \\ 
BM And\dotfill   & G8	& 2.03 & 3.02 & 510  & \nod & \nod & \nod & $>$-9   & 15,15,7      & 2 \\
CI Tau\dotfill   & K7	& 0.70 & 1.94 & 370  & \nod & \nod & 0.47 & -6.8    & 12,12,10     & 1 \\
CY Tau\dotfill   & M1	& 0.43 & 1.70 & 310  & 7.5  & 7.2  & 1.20 & -8.1    & 12,9,4,9     & 3 \\
DK Tau\dotfill   & K7	& 0.69 & 2.51 & 320  & 8.4  & 6.1  & 0.49 & -7.4    & 12,9,3,9     & 4 \\
DN Tau\dotfill   & M0	& 0.52 & 2.15 & 300  & 6.0  & 5.2  & 0.08 & -8.5    & 12,9,2,9     & 2 \\
DR Tau\dotfill   & K7	& 0.69 & 2.75 & 310  & 9.0  & 5.9  & 9.60 & -5.1    & 12,10,4,10   & 4 \\
DS Tau\dotfill   & K5	& 1.09 & 1.30 & 570  & \nod & \nod & 0.96 & -7.9    & 12,9,9       & 1 \\
FP Tau\dotfill   & M4	& 0.21 & 2.00 & 200  & \nod & \nod & 0.15 & -7.7    & 12,12,10     & 1 \\
GI Tau\dotfill   & K6	& 0.93 & 1.74 & 450  & 7.2  & 8.8  & 0.24 & -8.0    & 12,9,17,9    & 2 \\
GK Tau\dotfill   & K7	& 0.69 & 2.16 & 350  & 4.65 & 4.8  & 0.23 & -8.2    & 12,9,3,9     & 2 \\
HK Tau\dotfill   & M0.5 & 0.45 & 1.65 & 320  & \nod & \nod & 1.10 & -6.5    & 12,12,10     & 1 \\
LkCa 8\dotfill   & M0	& 0.53 & 1.48 & 370  & 3.25 & 5.1  & 0.15 & -9.1    & 12,9,3,9     & 2 \\
RW Aur B\dotfill & K5	& 0.96 & 1.09 & 580  & \nod & \nod & \nod & -8.8    & 19,19,19     & 1 \\
SU Aur\dotfill   & G2	& 2.02 & 3.27 & 490  & 1.7  & 2.3 & \nod & -8.0     & 12,12,6,8    & 1 \\
TW Hya\dotfill   & K7	& 0.75 & 1.04 & 520  & 2.80 & 7.3  & \nod & -9.3    & 18,18,13,14  & 6 \\
UY Aur\dotfill   & M0	& 0.54 & 1.30 & 400  & \nod & \nod & 0.40 & -7.6    & 11,11,11     & 4 \\
UZ Tau E\dotfill & M1	& 0.43 & 1.39 & 340  & \nod & \nod & 0.73 & -8.7    & 19,19,19     & 1 \\
UZ Tau W\dotfill & M2	& 0.33 & 1.88 & 260  & \nod & \nod & \nod & -8.0    & 11,11,11     & 1 \\
V836 Tau\dotfill & K7	& 0.71 & 1.43 & 440  & 7.0  & 9.6  & 0.05 & -8.2    & 12,12,16,10  & 2 \\
YY Ori\dotfill   & K7	& 0.68 & 3.00 & 290  & 7.58 & 4.8 & 1.80 & -5.5     & 10,10,1,10   & 1 \\
\enddata
\tablecomments{Col.~2: Spectral type; Col.~3: Stellar mass in $M_\odot$; Col.~4: Stellar radius in $R_\odot$; Col.~5: Stellar escape velocity in \kms, calculated from columns 3, 4; Col.~6: Rotation period in days; Col.~7: Corotation radius in $R_*$, calculated from columns 3, 4, 6; Col.~8: Median veiling at 5700~\AA\ from HEG; Col.~9: Logarithm of the mass accretion rate in \msunyr; Col.~10: References for the spectral type, stellar luminosity (to determine $M_*$ and $R_*$), rotation rate (where available), and mass accretion rate; Col.~11: Number of spectra acquired with NIRSPEC.}\tablerefs{(1) \citealt{ber96}; (2) \citealt{bou86}; (3) \citealt{bou93}; (4) \citealt{bou95}; (5) \citealt{bou07b}; (6) \citealt{dew03}; (7) \citealt{gue93}; (8) \citealt{gul00}; (9) \citealt{gul98}; (10) HEG; (11) \citealt{har03}; (12) \citealt{ken95}; (13) \citealt{law05}; (14) \citealt{muz00}; (15) \citealt{ros99}; (16) \citealt{ryd84}; (17) \citealt{vrb86};  (18) \citealt{web99}; (19) \citealt{whi01}.}
\end{deluxetable*}

As in the 2006 paper, the additional spectra were acquired with NIRSPEC on Keck II \citep{mcl98} using the N1 filter ($Y$ band), which covers the range 0.95 to 1.12~\micron\ at a resolution $R=25,000$ ($\Delta V=12$~\kms).  The echelle order of primary interest extends from 1.081 to 1.096~\micron\ and contains both \helium\ and \pgamma.  Spectra from the 2005-06 season were acquired by G. Blake (13 December 2005) and D. Stark (13 January 2006).  Those from November and December 2006 were obtained by L. Hillenbrand, W. Fischer, S. Edwards, and C. Sharon.  Finally, Hillenbrand obtained two additional spectra of TW Hya in December 2007.  Data reduction, including wavelength calibration and spatial rectification, extraction of one-dimensional spectra from the images, and removal of telluric emission and absorption features, is discussed in EFHK.  While we used an IRAF script to reduce the EFHK data, we used the IDL package REDSPEC by S. S. Kim, L. Prato, and I. McLean to reduce data acquired in Fall 2006 and later.  EFHK also describes the procedure for measuring photospheric lines to determine the 1-$\micron$ veiling $r_Y$, defined as the ratio of excess flux to photospheric flux near the \helium\ line \citep[see also][]{har89}.  After the veilings are determined, a non-accreting template that has been artifically veiled to match the observed CTTS is subtracted from each target spectrum.  This removes photospheric absorption lines from the \helium\ and \pgamma\ regions, which allows for a more accurate definition of the remaining structure in each of these two lines.

We augmented the spectral templates from those of EFHK, resulting in a reassessment of the 1-$\micron$ veiling for one object. Recent determinations of the spectral type of BM And in the V band range from G8 \citep{gue93} to K5 \citep{mor01}.  In EFHK we used an early K star to deveil the 2002 spectrum of BM And.  However, using our new grid of templates acquired in Fall 2006, we found that the G8 dwarf HD 75935 provides a better match to the photosphere of BM And.  Deveiling the 2002 spectrum of BM And with this template yields a veiling of 0.4, in contrast to the value of 0.1 reported in EFHK.  We adopt the more appropriate veilings for BM And in this work, $r_Y=0.4$ in 2002 and $r_Y=0.5$ in 2006. The veiling determinations for all the other objects from EFHK are unaffected by our extended grid of templates.

We use the stellar mass, radius, and rotation period to calculate the escape velocity and the star-disk corotation radius for each star, which are included in Table~\ref{t.sample} and will be used in later analysis of the accretion geometry.  We carefully surveyed the literature to acquire the most up-to-date estimates of spectral types and stellar luminosities, using \citet{ken95} and \citet{gul98} in most cases.  The luminosity of YY Ori (HEG) was updated to reflect the latest estimate of the distance to Orion \citep{men07}, which is 10\% less than the earlier value.  Spectral types were converted into effective temperatures using the scale from \citet{hil04}, and stellar radii follow directly from application of the Stefan-Boltzmann law to the effective temperatures and luminosities.  Stellar masses are then derived from the \citet{sie00} pre--main-sequence tracks, available online.  Since the escape velocity will be an important parameter in comparing observed to model profiles, we have given some thought to its accuracy.  With a dependency on $M/R$, the largest source of uncertainty in calculating the escape velocity is the uncertainty in $T_{\rm eff}$, since temperature strongly affects both the mass and radius determination, while luminosity only weakly influences the radius estimate.  We assess that the typical error in the escape velocity is $\sim20\%$.  For the 12 stars with rotation periods in the literature, we calculate corotation radii with a typical error of 20\%, provided the photometric period is equivalent to the rotation period.

Three of the 21 objects are known members of binary pairs resolved in our spectra where we have observed only the primary: DK Tau A, HK Tau A, and UY Aur A.  For another system, RW Aur, we have resolved spectra of RW Aur A and RW Aur B, but only the latter shows red absorption at \helium\ and thus qualifies as part of the sample for this study.  There is conflicting evidence in the literature on whether RW Aur B has a close companion at an angular separation of 0.12\arcsec\ with a K-band flux ratio of 0.024 \citep{ghe93,cor06}.  In our spectra we see the lines of only one object consistent with a K5 spectral type, and we call this RW Aur B.  An additional two objects are unresolved binaries: UZ Tau E and UZ Tau W.  For these, we also see lines from only one star and attribute the 1-$\micron$ continuum and line profiles to the primary.

\section{EMPIRICAL RESULTS\label{s.obs}}

In this study we concentrate on the red absorption seen at \helium\ as a probe of the accreting gas, ignoring the blue absorptions that arise from disk and stellar winds (KEF). For each spectrum of the 21 stars that show subcontinuum redshifted \helium\ absorption at least once, Table~\ref{t.redabs} lists the HJD of observation, the 1-$\micron$ veiling, and measurements of the red absorption. In this section we first compare the veilings of the stars in this study to those from the ensemble of 38 CTTS in EFHK, and then we present the profiles and kinematic data for the {\it reference sample}, consisting of the single observation of each star with the deepest red absorption (identified with an asterisk in Table~\ref{t.redabs}). Next we demonstrate that the propensity for \helium\ to absorb all impinging 1-$\micron$ photons provides a means of estimating the origination radius in the disk for infalling gas and the filling factor of accreting material immediately before the accretion shock. We conclude the section with a discussion of profile and veiling variability.

\begin{deluxetable*}{lcccccccc}
\tablecaption{Veilings and Measurements of \helium\ Subcontinuum Red Absorption\label{t.redabs}}
\tablewidth{6in}
\tablehead{& & & \colhead{$W_\lambda$} & \colhead{$D_{\rm max}$} & \colhead{$V_C$} & \colhead{FWQM} & \colhead{$V_{\rm{blue}}$} & \colhead{\vred}\\ \colhead{Object} & \colhead{HJD} & \colhead{$r_Y$} & \colhead{(\AA)} & \colhead{(\%)} & \colhead{(\kms)} & \colhead{(\kms)} & \colhead{(\kms)} & \colhead{(\kms)}\\ \colhead{(1)} & \colhead{(2)} & \colhead{(3)} & \colhead{(4)} & \colhead{(5)} & \colhead{(6)} & \colhead{(7)} & \colhead{(8)} & \colhead{(9)}}
\startdata
AA Tau\dotfill   &  605.0  & 0.2  & 0.5 & 14   & 185  & 120  & 110  & 250  \\
                 &  606.9  & 0.1  & 1.9 & 42   & 110  & 170  & 40   & 250  \\
                 & 1718.0* & 0.0  & 4.5 & 61   & 90   & 310  & -40  & 310  \\
                 & 2069.0  & 0.1  & 0.9 & 24   & 50   & 70   & 0\tablenotemark{a}    & 80   \\
BM And\dotfill   &  604.8  & 0.4  & 2.3 & 28   & 115  & 260  & -20  & 290  \\
                 & 2068.7* & 0.5  & 2.9 & 40   & 90   & 270  & -40  & 280  \\
CI Tau\dotfill   &  605.9  & 0.2  & 1.3 & 17   & 140  & 290  & 10   & 310  \\
CY Tau\dotfill   &  606.8  & 0.1  & 1.0 & 27   & 140  & 120  & 80   & 230  \\
                 & 1718.0* & 0.0  & 1.1 & 37   & 140  & 100  & 80   & 240  \\
                 & 2068.8  & 0.2  & 0.0 & \nod & \nod & \nod & \nod & \nod \\
DK Tau\dotfill   &  604.9  & 0.5  & 2.1 & 28   & 145  & 290  & 20   & 330  \\
                 &  606.9  & 0.5  & 1.9 & 37   &  80  & 240  & 0\tablenotemark{a}    & 280  \\
                 & 1748.9* & 0.0  & 3.1 & 40   & 150  & 320  & -20  & 340  \\
                 & 2068.9  & 0.4  & 1.9 & 34   & 95   & 290  & 0\tablenotemark{a}    & 310  \\
DN Tau\dotfill   &  606.0  & 0.0  & 1.3 & 30   & 145  & 170  & 60   & 250  \\
                 & 1718.0* & 0.0  & 1.2 & 33   & 150  & 140  & 70   & 260  \\ 
DR Tau\dotfill   &  605.0  & 2.0  & 0.0 & \nod & \nod & \nod & \nod & \nod \\
                 &  606.0  & 2.0  & 0.0 & \nod & \nod & \nod & \nod & \nod \\
                 &  606.9* & 2.0  & 0.7 & 14   & 235  & 160  & 150  & 320  \\
                 & 2069.1  & 3.5  & 0.0 & \nod & \nod & \nod & \nod & \nod \\
DS Tau\dotfill   &  605.9  & 0.4  & 1.1 & 18   & 205  & 240  & 90   & 340  \\
FP Tau\dotfill   &  605.0  & 0.1  & 0.5 & 17   & 50   & 120  & 0\tablenotemark{a}    & 120  \\
GI Tau\dotfill   &  606.0  & 0.1  & 3.1 & 47   & 160  & 230  & 50   & 330  \\
                 & 2069.8* & 0.0  & 3.3 & 52   & 180  & 240  & 50   & 350  \\
GK Tau\dotfill   &  606.0  & 0.3  & 0.0 & \nod & \nod & \nod & \nod & \nod \\
	         & 2069.8* & 0.1  & 0.5 & 11   & 160  & 140  & 50   & 220  \\
HK Tau\dotfill   &  606.1  & 0.4  & 0.5 & 17   & 75   & 100  & 30   & 140  \\
LkCa 8\dotfill   &  604.9* & 0.05 & 1.4 & 32   & 160  & 160  & 70   & 280  \\
                 & 2068.9  & 0.1  & 1.2 & 24   & 125  & 190  & 40   & 250  \\
RW Aur B\dotfill &  605.1  & 0.1  & 2.8 & 43   & 160  & 230  & 50   & 330  \\
SU Aur\dotfill   &  607.0  & 0.0  & 1.6 & 35   & 50   & 180  & -50  & 150  \\
TW Hya\dotfill   &  605.2  & 0.0  & 0.0 & \nod & \nod & \nod & \nod & \nod \\
                 &  606.1  & 0.0  & 0.0 & \nod & \nod & \nod & \nod & \nod \\
                 & 1718.1* & 0.1  & 1.4 & 32   & 245  & 170  & 170  & 370  \\
                 & 2069.1  & 0.1  & 0.9 & 17   & 255  & 170  & 170  & 350  \\
                 & 2452.1  & 0.0  & 0.8 & 17   & 230  & 190  & 150  & 350  \\
                 & 2453.1  & 0.1  & 0.7 & 14   & 240  & 190  & 160  & 330  \\
UY Aur\dotfill   &  605.0  & 0.4  & 0.4 & 14   & 160  & 110  & 110  & 220  \\
                 &  607.0  & 0.4  & 0.4 & 12   & 160  & 110  & 100  & 220  \\
                 & 1718.1* & 0.2  & 0.7 & 20   & 160  & 130  & 90   & 240  \\
                 & 2069.9  & 0.3  & 0.5 & 13   & 160  & 120  & 90   & 230  \\
UZ Tau E\dotfill &  605.9  & 0.3  & 0.2 &  8   & 185  & 60   & 150  & 210  \\
UZ Tau W\dotfill &  605.9  & 0.1  & 0.6 & 18   & 75   & 140  & 20   & 170  \\
V836 Tau\dotfill &  606.0* & 0.0  & 1.7 & 35   & 160  & 170  & 80   & 300  \\
                 & 1749.0  & 0.0  & 0.0 & \nod & \nod & \nod & \nod & \nod \\
YY Ori\dotfill   &  607.1  & 0.4  & 2.1 & 37   & 225  & 210  & 110  & 390  \\
\enddata
\tablecomments{Col.~2: Heliocentric Julian Date (2,452,000 +); for multiple observations an asterisk indicates membership in the reference sample; Col.~3: Veiling at one micron; Col.~4: Equivalent width of red absorption below the continuum; Col.~5: Percentage of the continuum absorbed at the deepest point of the profile; Col.~6: Centroid of the red absorption; Col.~7: Width at one quarter of red absorption minimum; Col.~8: Minimum velocity of red absorption; Col.~9: Maximum velocity of red absorption.}
\tablenotetext{a}{The true minimum velocity is obscured by central
absorption; we assume $V_{\rm blue}=0$.}
\end{deluxetable*}

\subsection{Veiling and Redshifted Absorption}

\begin{figure}
\epsscale{1.2}
\plotone{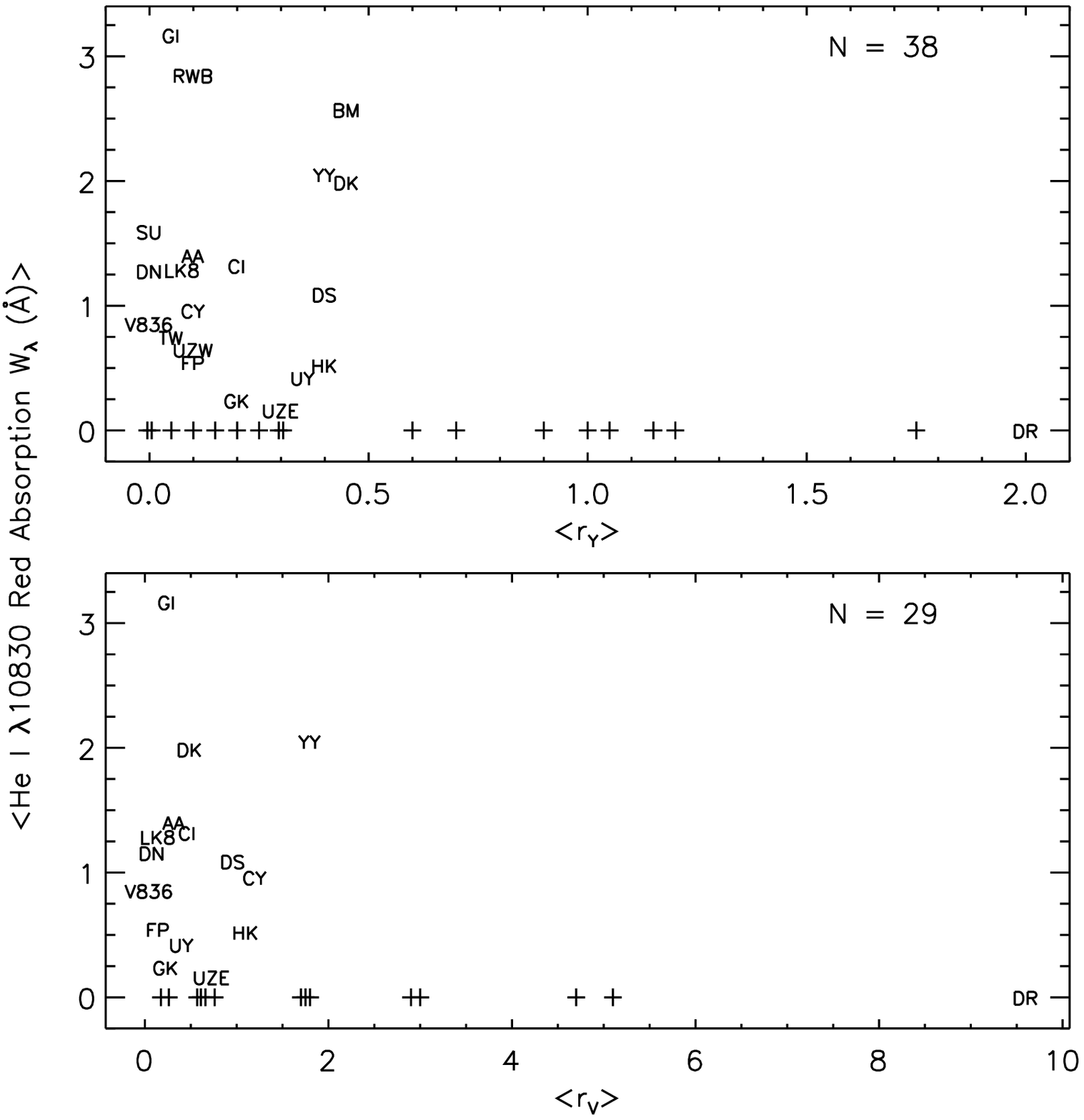}
\figcaption{ Equivalent width of red absorption at \helium\ versus veiling for CTTS from EFHK.  The top panel shows the relation for the simultaneously measured 1-$\micron$ veiling $r_Y$ for all 38 CTTS, using averages  for stars with multiple observations.   The 21 stars featured here show helium red absorption in at least one observation and are labeled with abbreviations of their names. The 17 that have not been observed to show red absorption are identified with plus signs. The bottom panel plots the same equivalent width data versus the average of optical veiling measurements ($\lambda=5700$~\AA) from HEG, obtained a decade before the NIRSPEC campaign, which exist for 29 of the EFHK stars. In this and future scatter plots, points that would otherwise overlap are slightly offset for clarity. \label{f.veil}}
\epsscale{1}
\end{figure}

Our additional observations beyond those in EFHK confirm the result reported therein, that subcontinuum redshifted absorption is more prevalent in CTTS with low veiling. We illustrate this in Figure~\ref{f.veil}, where the equivalent width of the redshifted \helium\ absorption below the continuum is plotted against both the simultaneous 1-$\micron$ veiling $r_Y$ and the non-simultaneous optical veiling $r_V$ for the 38 CTTS in EFHK. The 21 CTTS that are the focus of this study, showing redshifted absorption at least once among 46 spectra, are each identified by name. The remaining 17 CTTS that have not yet been seen to show redshifted absorption among 35 spectra acquired to date appear as symbols (but can be identified from EFHK).  All points in the figure are averages for objects with multiple observations taken between 2001-2007.

We have included the optical veiling measurements in Figure~\ref{f.veil} because it is the excess emission at optical and shorter wavelengths that is associated with luminosity from accretion shocks and is the basis for deriving disk accretion rates. Note that the range of veilings is different at each of the two wavelengths, with maxima of $r_V=9.6$ and $r_Y=2$, and that although all CTTS show detectable veiling in the optical (HEG), 7/38 have no detectable veiling at 1 $\micron$.  (All the CTTS  with $r_Y=0$ do show \helium\ and \pgamma\ in emission, differentiating them from WTTS.)  Despite these differences, objects with low $r_V$ have low $r_Y$, and objects with high $r_V$ have high $r_Y$, clarifying that the 1-$\micron$ veiling is a rough proxy for optical/UV veiling and disk accretion rate, and that variations in the disk accretion rate are relatively modest over timescales of a decade.   The proportionality between $r_Y$ and disk accretion rate is further corroborated by the excellent correlation between $r_Y$ and the equivalent width of \pgamma\ emission (EFHK), assuming that the equivalent width of \pgamma, like that of P$\beta$, is correlated with accretion rate \citep{muz98,fol01,nat04}.

The prime message from Figure~\ref{f.veil} is that when the veiling is high, $r_Y>0.5$ or $r_V>2$, red absorption at \helium\ is rare. Although the number of observations of each of the 38 stars ranges from 1 to 6, we note that out of a total of 25 observations of the 9 objects with $r_Y>0.5$, only once, in one of four observations of the highest-veiling object DR Tau, did a weak redshifted absorption appear.  In contrast, out of the 56 total spectra of the 29 objects with $r_Y\le0.5$ or $r_V\le2$, redshifted absorption is detected in 37 spectra. Even with our non-uniform sampling of individual objects, it is clear that the frequency of redshifted absorption in CTTS with the highest veilings, seen in only 4\% (1/25) of the total spectra of 9 objects, is significantly lower than that in CTTS with more modest veilings, where red absorption is seen in 66\% (37/56) of the total spectra of 29 objects. Spectral variability for these objects will be discussed in Section~3.4.

\subsection{Line Profiles and Subcontinuum Absorption}

\begin{figure*}
\plotone{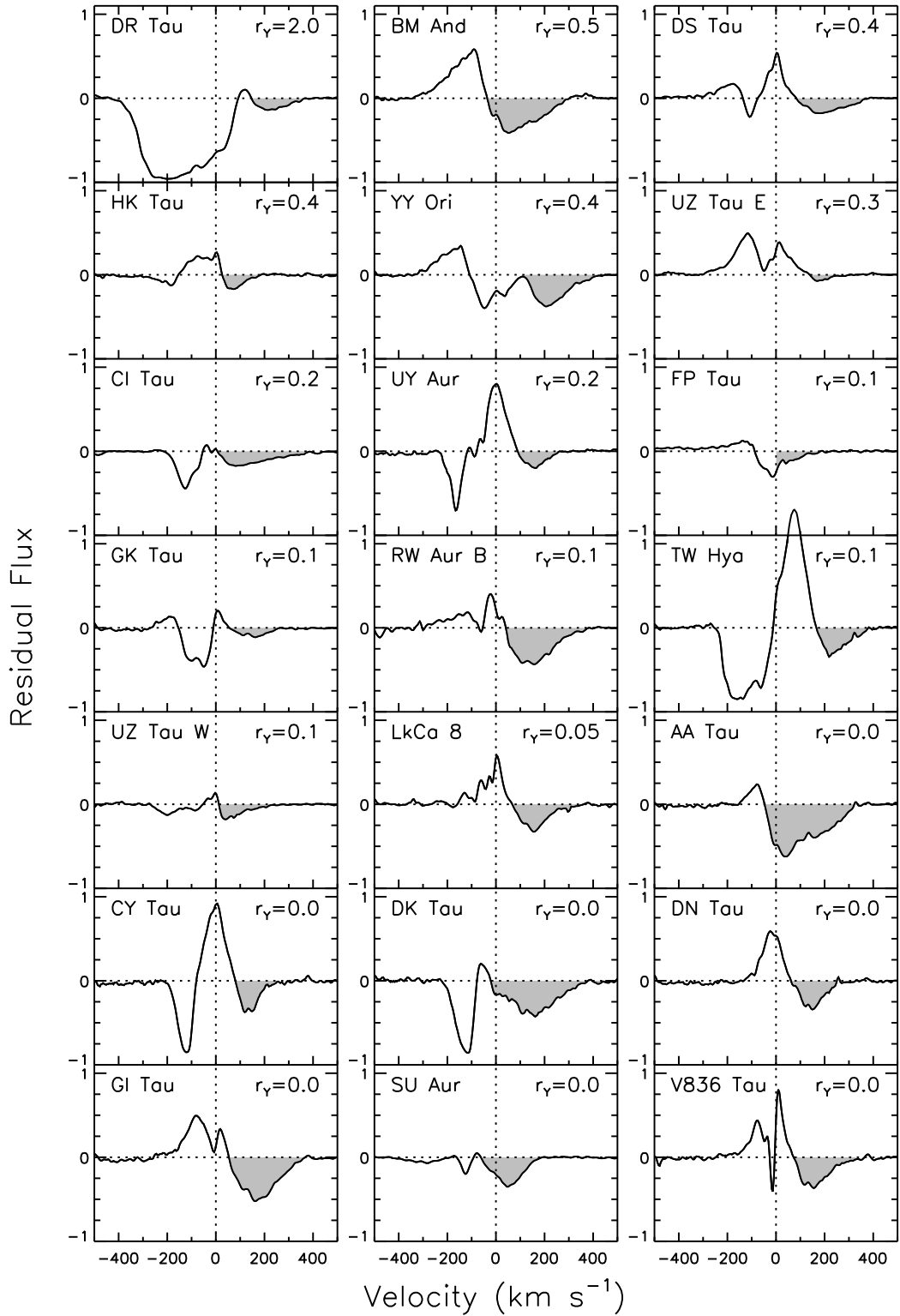}
\figcaption{The reference sample of residual \helium\ profiles for the 21 CTTS with subcontinuum redshifted absorption (shaded), ordered by decreasing 1-$\micron$ veiling $r_Y$.  Since the veiled photospheric contribution has been subtracted, the continuum corresponds to zero on the flux axis, and total absorption of the continuum corresponds to -1.  Velocities are relative to the stellar photosphere, and the spectra are plotted with three-pixel binning.\label{f.redabs}}
\end{figure*}

\helium\ profiles for the 21 CTTS that have shown subcontinuum redshifted absorption at least once are presented in Figure~\ref{f.redabs}. This set of profiles is for the reference sample, and the profiles are ordered by their simultaneous 1-$\micron$ veiling. The part of the profile we identify as the red absorption component is delineated in Figure~\ref{f.redabs} by shading. As noted above, the reference sample contains the profile with the deepest red absorption at \helium\ for each star (in contrast to the reference sample from EFHK that emphasized blueshifted absorption from winds).   The full set of profiles observed for the 12 stars with multiple spectra appears in Section~\ref{s.var} where we discuss variability. The reference sample will be used in all subsequent analysis unless we are explicitly considering profile or veiling variations.  
 
\begin{figure*}
\plotone{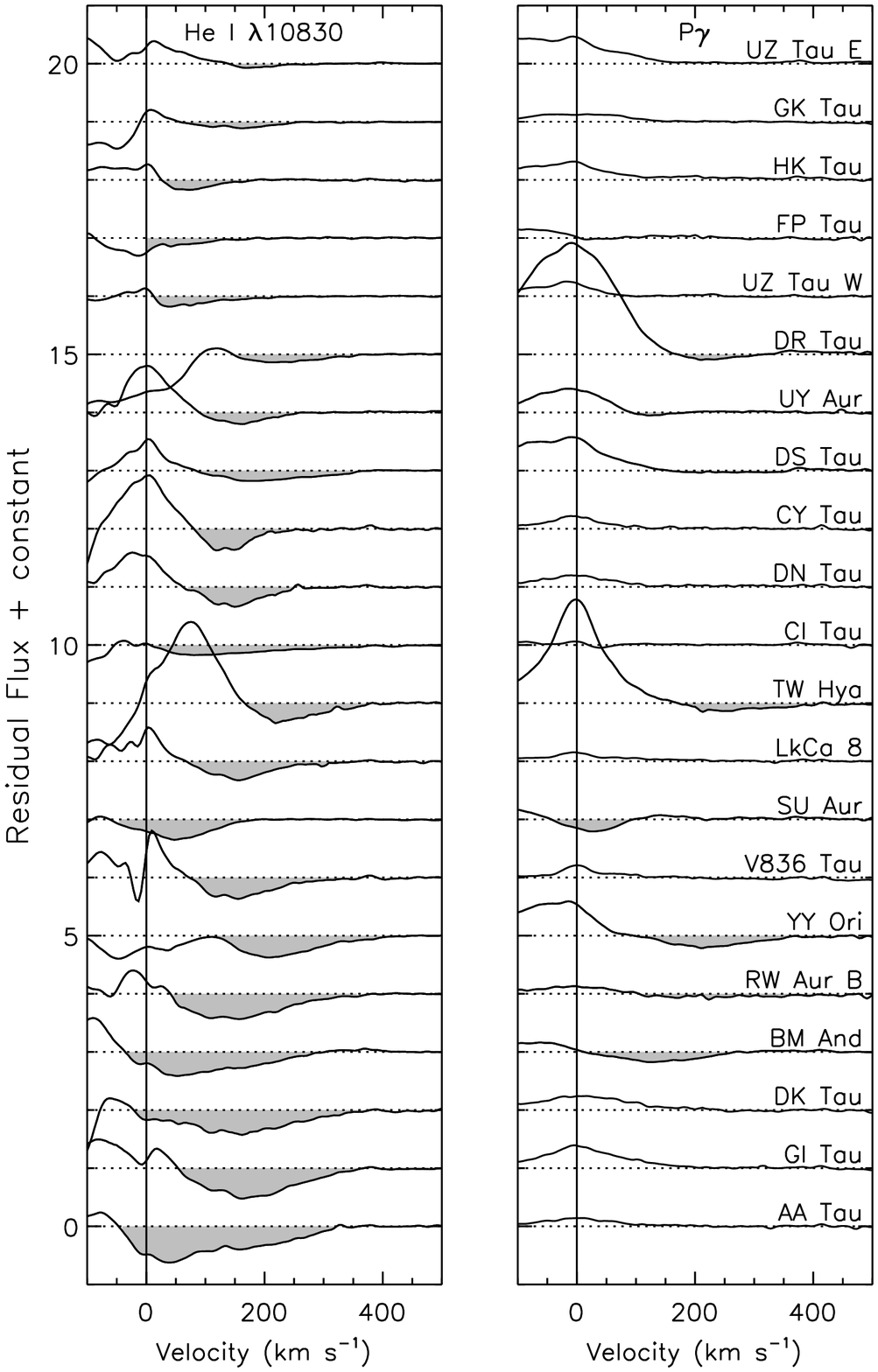}
\figcaption{Comparison of the red half of \helium\ (left) and \pgamma\ (right) profiles from the reference sample, arranged top to bottom in order of increasing \helium\ red absorption equivalent width. Subcontinuum absorption is shaded in both lines.\label{f.width}}
\end{figure*}

The uniqueness of the \helium\ line in the study of accreting gas is immediately seen by comparing it to the \pgamma\ profile found in the same NIRSPEC order. This comparison is made in Figure~\ref{f.width}, which zooms in on the red side of the profile for each of the 21 objects in the reference sample, sorted now by the equivalent width of the \helium\ red absorption.   (We have ignored the blue half of the line in order to draw attention to the red absorption; full \pgamma\ profiles can be found in EFHK.) Only 5 of 21 (24\%) stars show red absorption at both \helium\ and \pgamma, and when seen, it is considerably weaker at \pgamma. Specifically, the maximum depth of red absorption seen at \pgamma\ is 21\% of the continuum, compared to 61\% for $\lambda$10830, and the maximum equivalent width is 1.1~\AA, versus 4.5~\AA\ at $\lambda$10830.  Surprisingly, the 3 stars with the strongest helium absorption (AA Tau, GI Tau, and DK Tau) show no absorption at \pgamma, while the 3 stars with the strongest \pgamma\ absorption (TW Hya, BM And, and YY Ori) have intermediate helium absorptions, although their \pgamma\ absorptions do share similar velocity structures with their helium absorptions.  In models for H line formation in magnetospheric accretion scenarios, inverse P Cygni absorption is seen only when the accretion rate is favorable and the line of sight is directed toward the hot continuum in the accretion shock.  In contrast, our data indicate a much wider range of formation conditions for red absorption at \helium, which offers a unique probe of the infalling gas projected in front of the stellar surface by absorbing continuum photons from both the star and the accretion shock.

\begin{figure}
\epsscale{1.2}
\plotone{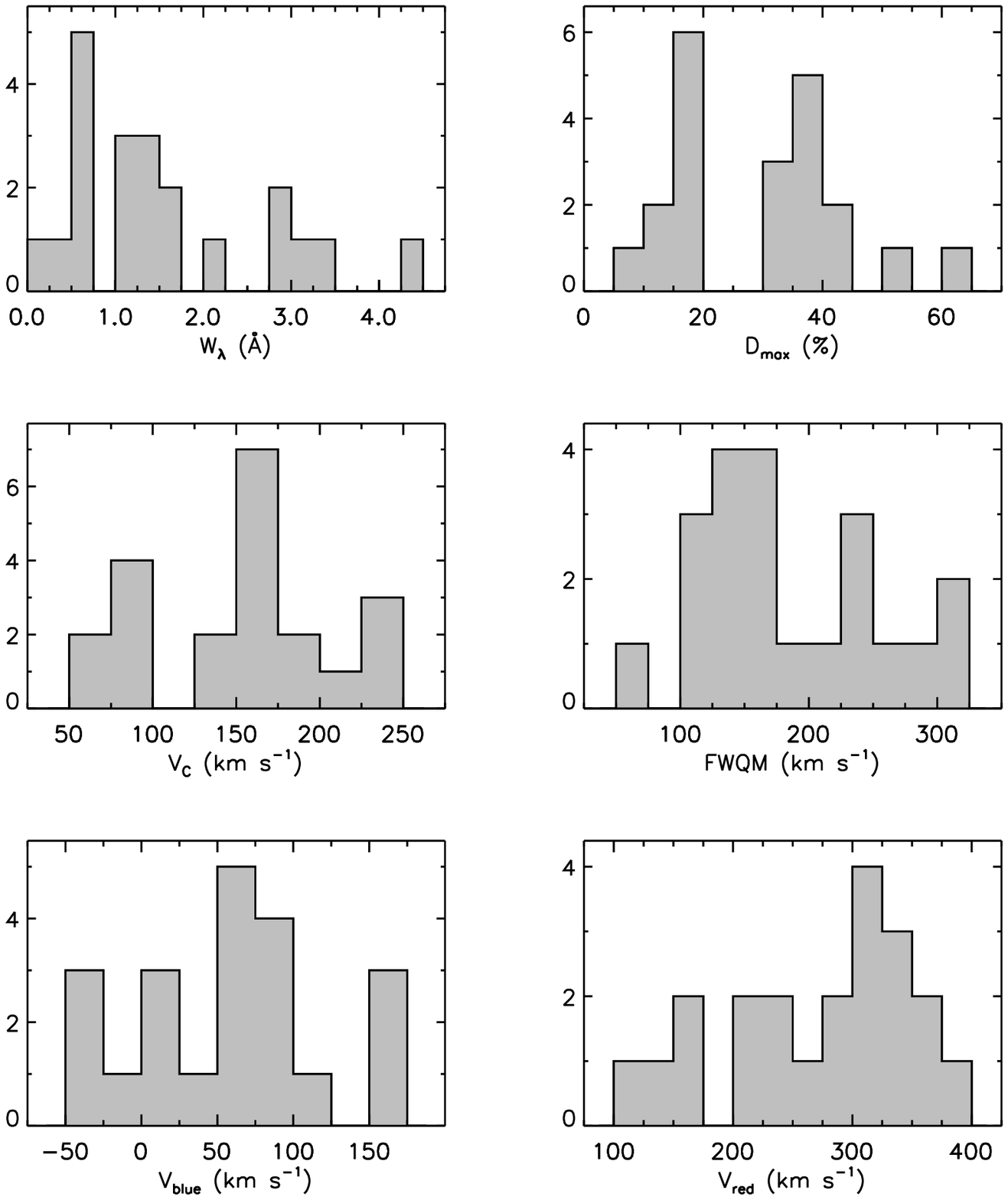}
\figcaption{Histograms of red absorption profile measurements for the reference sample.\label{f.hist}}
\epsscale{1}
\end{figure}

Measured parameters of the \helium\ red absorption, i.e., the section of the profile shaded in Figure~\ref{f.redabs}, are listed for each observation in Table~\ref{t.redabs} with an asterisk identifying the spectrum in the reference sample for stars with multiple observations. Parameters include the equivalent width $W_\lambda$, the depth of maximum penetration into the continuum $D_{\rm max}$, the centroid velocity $V_C$, and the width measured at one quarter of the absorption minimum, FWQM.  We also tabulate velocities at the blueward and redward edges of the absorption, $V_{\rm blue}$ and \vred.  In most cases, $V_{\rm blue}$ is easily identified as the location where emission sharply transitions to red absorption.  On the other hand, the gradual return to the continuum at the high-velocity end makes \vred\ less straightforward to measure.  In order to have a uniform definition for all stars, we conservatively define \vred\ as the velocity where the absorption reaches 95\% of the continuum level, with the consequence that it is somewhat smaller than the extreme infall velocity. Histograms illustrating the diversity of these parameters appear in Figure~\ref{f.hist}.  The equivalent widths range from 0.2 to 4.5~\AA, maximum penetrations into the continuum range from 8\% to 61\%, centroids range from 50 to 255~\kms, and FWQM range from 60 to 320~\kms.  In many stars the absorptions begin near the stellar rest velocity, so the FWQM reflects the true width of the absorbing velocities. In others $V_{\rm blue}$ is well redward of the rest velocity (e.g., DR Tau and TW Hya) due to the presence of helium emission that is likely from another region, such as the wind, that is filling in the red absorption and reducing its magnitude.

\subsection{Maximum Infall Velocities\label{s.rred}}

Without assuming any particular infall geometry, we can estimate the outer extent of an accretion flow by comparing the most redward velocity in an absorption profile with the stellar escape velocity.  A particle undergoing ballistic infall from a distance $R$ toward a star of mass $M_*$ and radius $R_*$ has a free-fall speed at a distance $r$ of \begin{eqnarray}v_{ff}&=&\left[\frac{2GM_*}{R_*}\left(\frac{R_*}{r}-\frac{R_*}{R}\right)\right]^{1/2}\nonumber\\ &=&V_{\rm esc}\left(\frac{R_*}{r}-\frac{R_*}{R}\right)^{1/2}\label{e.speed}\end{eqnarray} where \vesc\ is the escape velocity from the surface of the star.  The largest infall velocity achieved in the funnel flow, immediately before impact,  is thus set by the maximum distance where infalling gas leaves the disk, $R_{\rm max}$, i.e., $V_{\rm max}=V_{\rm esc}\left(1-R_*/R_{\rm max}\right)^{1/2}$.  We can then use the velocity of the red edge of the \helium\ absorption, \vred, as an indicator for $V_{\rm max}$ and hence determine $R_{\rm max}$.  The depth of the absorption near \vred\ will then indicate the filling factor of infalling gas immediately before it impacts the photosphere.  However, because of projection effects and the conservative assumption for measuring  \vred, this gives a lower limit to the actual $V_{\rm max}$, and thus the corresponding inferred maximum distance for infall, $R_{\rm red}$, will be a lower limit to the actual value of $R_{\rm max}$. Thus $V_{\rm max}\ge V_{\rm red}$ and with $R_{\rm max}$ in units of $R_*$, we have \begin{equation}R_{\rm max}\ge\left(1-V_{\rm red}^2/V_{\rm esc}^2\right)^{-1}\equiv R_{\rm red}.\label{e.rred}\end{equation}

\begin{figure}
\epsscale{1.2}
\plotone{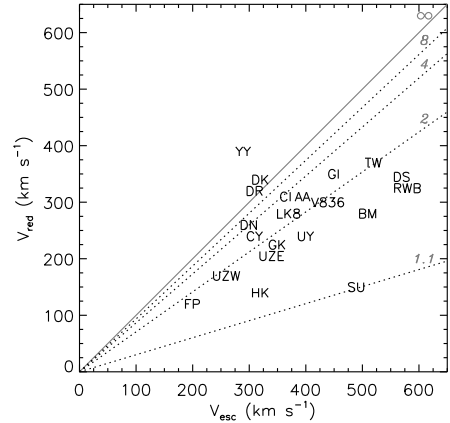}
\figcaption{Velocity at the red edge of the \helium\ absorption (\vred) versus the stellar escape velocity (\vesc) for the 21 profiles of the reference sample.  The solid line represents $V_{\rm red}=V_{\rm esc}$, while the dotted lines represent the ratios of \vred\ to \vesc\ for ballistic infall from 8, 4, 2, and 1.1~$R_*$, assuming $V_{\rm red}=V_{\rm max}$.\label{f.rmax}}
\epsscale{1}
\end{figure}

In Table~\ref{t.radius} we list \vred, \vesc, and their ratio, followed by the implied $R_{\rm red}$.  Figure~\ref{f.rmax} shows the locations of the 21 observations of the reference sample in the (\vred, \vesc) space as well as dotted lines corresponding to $V_{\rm red}/V_{\rm esc}$ of 0.94, 0.87, 0.71, and 0.30, or $R_{\rm red}=8$, 4, 2, and 1.1~$R_*$ respectively.  The average value of $R_{\rm red}$ is 2.9 $R_*$, and the median is 1.9 $R_*$, where we adopt $R_{\rm red}\ge8R_*$ for the 3 stars with $V_{\rm red}>V_{\rm esc}$.  Of these 3 outliers, YY Ori has $V_{\rm red}/V_{\rm esc}=1.4$, which indicates an error in the stellar parameters, while the other two, DR Tau and DK Tau, have \vred\ slightly larger than \vesc\ but within the estimated 20\% uncertainty.  If we used a less conservative estimate for \vred\ as the outermost redward velocity (see Section 3.2), at a penetration depth of 2\% rather than 5\% of the continuum, then the median $R_{\rm red}$ increases to 2.9 $R_*$. 

\begin{deluxetable*}{lcccccc}
\tablecaption{Maximum Infall Distances and Corotation Radii\label{t.radius}}
\tablewidth{6in}
\tablehead{\colhead{Object} & \colhead{\vred} & \colhead{\vesc} & \colhead{$V_{\rm red}/V_{\rm esc}$} & \colhead{$R_{\rm red}$\tablenotemark{a}} & \colhead{$R_{\rm co}$\tablenotemark{b}} & \colhead{$R_{\rm red}/R_{\rm co}$\tablenotemark{c}} \\ \colhead{(1)} & \colhead{(2)} & \colhead{(3)} & \colhead{(4)} & \colhead{(5)} & \colhead{(6)} & \colhead{(7)}}
\startdata
AA Tau\dotfill   & 310 & 390  & 0.79 &  2.7      & 8.7   & 0.3          \\
BM And\dotfill   & 280 & 510  & 0.55 &  1.4      & [6.3] & [0.2]        \\
CI Tau\dotfill   & 310 & 370  & 0.84 &  3.4      & [6.3] & [0.5]        \\
CY Tau\dotfill   & 240 & 310  & 0.77 &  2.5      & 7.2   & 0.4          \\
DK Tau\dotfill   & 340 & 320  & 1.06 &  [$\ge$8] & 6.1   & [$\ge$1.3]   \\
DN Tau\dotfill   & 260 & 300  & 0.87 &  4.0      & 5.2   & 0.8          \\
DR Tau\dotfill   & 320 & 310  & 1.03 &  [$\ge$8] & 5.9   & [$\ge$1.4]   \\
DS Tau\dotfill   & 340 & 570  & 0.60 &  1.6      & [6.3] & [0.3]        \\
FP Tau\dotfill   & 120 & 200  & 0.60 &  1.6      & [6.3] & [0.3]        \\
GI Tau\dotfill   & 350 & 450  & 0.78 &  2.5      & 8.8   & 0.3          \\
GK Tau\dotfill   & 220 & 350  & 0.63 &  1.7      & 4.8   & 0.4          \\
HK Tau\dotfill   & 140 & 320  & 0.44 &  1.2      & [6.3] & [0.2]        \\
LkCa 8\dotfill   & 280 & 370  & 0.76 &  2.3      & 5.1   & 0.5          \\
RW Aur B\dotfill & 330 & 580  & 0.57 &  1.5      & [6.3] & [0.2]        \\
SU Aur\dotfill   & 150 & 490  & 0.31 &  1.1      & 2.3   & 0.5          \\
TW Hya\dotfill   & 370 & 520  & 0.71 &  2.0      & 7.3   & 0.3          \\
UY Aur\dotfill   & 240 & 400  & 0.60 &  1.6      & [6.3] & [0.3]        \\
UZ Tau E\dotfill & 210 & 340  & 0.62 &  1.6      & [6.3] & [0.3]        \\
UZ Tau W\dotfill & 170 & 260  & 0.65 &  1.7      & [6.3] & [0.3]        \\
V836 Tau\dotfill & 300 & 440  & 0.68 &  1.9      & 9.6   & 0.2          \\
YY Ori\dotfill   & 390 & 290  & 1.34 &  [$\ge$8] & 4.8   & [$\ge$1.7] \\
\enddata
\tablecomments{For the reference sample only.  Col.~2: Maximum velocity of \helium\ red absorption (\kms); Col.~3: Escape velocity from the stellar surface (\kms); Col.~5: Lower limit to maximum distance of infalling material ($R_*$); Col.~6: Corotation radius ($R_*$).}
\tablenotetext{a}{Brackets indicate an assumed $R_{\rm red}\ge8~R_*$ since $V_{\rm red}>V_{\rm esc}$.}

\tablenotetext{b}{Brackets indicate that, since $P_{\rm rot}$ and thus $R_{\rm co}$ are unknown, $R_{\rm co}$ is set to the mean of the known values.}

\tablenotetext{c}{Brackets indicate uncertainty due to one of the preceding conditions.}
\end{deluxetable*}

\begin{figure}
\epsscale{1.2}
\plotone{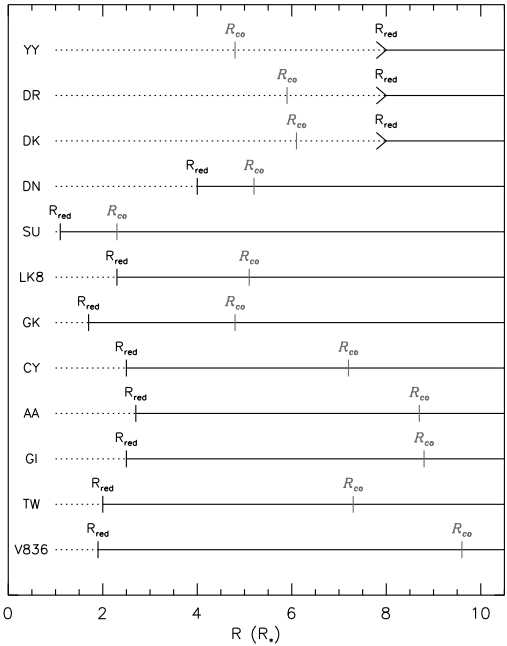}
\figcaption{Graphical comparison between $R_{\rm red}$, the {\em minimum} distance from the star of the outermost edge of the flow estimated from \vred, and $R_{\rm co}$, the star-disk corotation radius, for the twelve stars with known $P_{\rm rot}$. Sort is by $R_{\rm red}/R_{\rm co}$, which decreases from top to bottom.  Lower limits to $R_{\rm red}$ are indicated by greater-than symbols.\label{f.rcoro}}
\epsscale{1}
\end{figure}

 In Figure~\ref{f.rcoro} we compare $R_{\rm red}$ to $R_{\rm co}$ for the 12 stars with published rotation periods (listed in Table~\ref{t.sample}).  For this group of stars, the median ratio of $R_{\rm red}$ to $R_{\rm co}$ is 0.4, which increases to 0.5 for the less conservative estimate of the maximum redward velocity and can increase further when projection effects are considered.  This is well inside the corotation radius for most stars, but as will be apparent from model profiles in Section 4, for some viewing angles projection effects can result in a significant underestimate of $R_{\rm max}$ as determined from $R_{\rm red}$. Three stars show $R_{\rm red}/R_{\rm co}>1.3$ (YY Ori, DR Tau, and DK Tau), indicating that infall originates close to or possibly outside corotation, unless the error in $R_{\rm co}$ is larger than the typical 20\% uncertainty.

The projected area of the accretion flow immediately above the accretion shock can be equated to the depth of the absorption at the highest velocity in the red absorption profile. This estimate will be most reliable for objects where both (1) \vred\ is near \vesc, indicating that projection effects are not significantly altering our determination of the velocity just before impact, and (2) the 1-$\micron$ veiling is near zero, indicating that the red absorption is solely due to scattering of stellar photons so the absorption depth at each velocity is the minimum percentage of the stellar surface that is obscured at that velocity.  In the reference sample, three stars with $V_{\rm red}\ge0.8~V_{\rm esc}$ and no 1-$\micron$ veiling have significant absorption depths at 0.9~\vred: AA Tau (14\%), DK Tau (10\%), and GI Tau (10\%).  Their implied projected areas of material moving close to the escape velocity are an order of magnitude larger than the accretion shock filling factors estimated from their optical continuum excesses (CG).  Although accretion flows do ``funnel'' into narrow columns as they arrive at magnetic footpoints on the stellar surface, the coverage fraction of the flow in typical dipole flow geometries diminishes by less than 50\% from $\sim1.2~R_*$ to $R_*$ as the speed increases from $\sim0.9$~\vred\ to \vred, not by an order of magnitude. This hints that in at least several stars, a conventional dipolar accretion flow will have difficulty reconciling small shock filling factors with deep, broad red absorptions. \\

\subsection{Variability\label{s.var}}

\begin{figure*}
\plotone{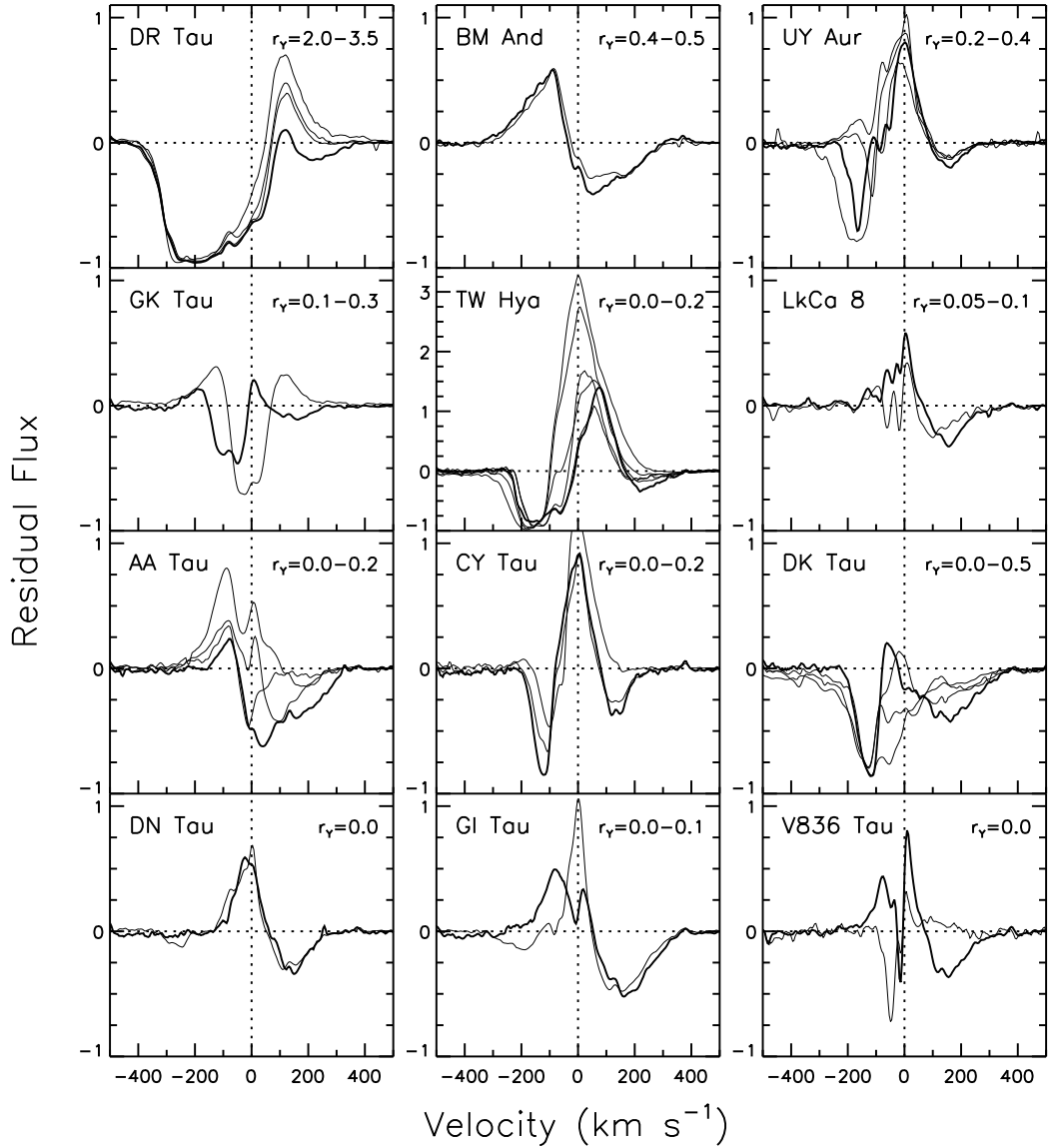}
\figcaption{Residual \helium\ profiles of the twelve CTTS that were observed more than once and that show subcontinuum redshifted absorption in at least one observation.  The reference sample spectra are shown with heavier lines, and the range of observed 1-$\micron$ veilings appears in each box.\label{f.multi}}
\end{figure*}

The \helium\ profiles for the 12 CTTS in this study with multiple spectra are shown in Figure~\ref{f.multi}, where for each star the full set of observed profiles is superposed and the range of simultaneous veilings is indicated. Since the time intervals are randomly distributed, ranging from days to years, only very general statements about variability can be made. Three categories of variability are seen: (1) five objects always show redshifted absorption with little variation in the absorption morphology  (BM And, UY Aur, LkCa 8, DN Tau, and GI Tau); (2) two objects always show red absorption, but the profile morphology changes dramatically (AA Tau and DK Tau); and (3) five objects have no redshifted absorption at one epoch, but do show it at another epoch (DR Tau, GK Tau, TW Hya, CY Tau, and V836 Tau). 

\begin{figure}
\epsscale{1.2}
\plotone{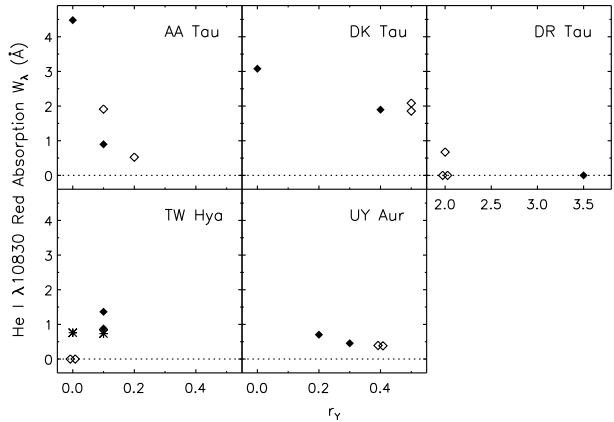}
\figcaption{Equivalent width of red absorption at \helium\ versus the 1-$\micron$ veiling for stars with at least four observations and at least one helium profile with subcontinuum red absorption.  Observations on contiguous days are represented by open points for 2002 and asterisks for 2007.\label{f.veilvar}}
\epsscale{1}
\end{figure}

The 5 stars observed at least four times (AA Tau, DK Tau, DR Tau, TW Hya and UY Aur) can be examined to see if there is a relation between veiling and the \helium\ red absorption.  The red absorption equivalent width is plotted against $r_Y$ for each observation of these five stars in Figure~\ref{f.veilvar}.  For the star with little change in the morphology of its red absorption (UY Aur), the veiling varies by a factor of 2. For the two stars where redshifted absorption is always present but changes dramatically (AA Tau and DK Tau), the absorption is strongest when the veiling is lowest (i.e., not detected). For the two stars where redshifted absorptions come and go (DR Tau and TW Hya), there is no relation between veiling and the strength of the absorption.

Each of the 5 stars with at least 4 observations was observed on at least two nights of a three-night run in 2002, providing a look at short-term variability and the possible role of rotation.  Data points from this run appear in Figure~\ref{f.veilvar} as open symbols, and the pair of asterisks for TW Hya are from a second set of two consecutive nights 5 years hence in 2007.  The only objects to show much variation over a time scale of days are DR Tau, which showed weak red absorption only on the last of three consecutive nights in 2002, and AA Tau, which we now discuss further.

The variability of AA Tau at optical wavelengths has been thoroughly examined in the context of rotational modulation from a misaligned magnetosphere interacting with the inner disk \citep{bou99,bou03,bou07b}.  Briefly, the system is close to edge-on, with an inclination angle of 75$^\circ$, and the rotation period is 8.22 days.  Phase zero corresponds to the epoch of maximum V-band flux, while phase 0.5 is characterized by a reduction in V-band flux due to occultation of the star by a warped disk.  Accretion diagnostics are strongest near phase 0.5, with redshifted absorption appearing at H$\alpha$ and H$\beta$ between phase 0.39 and phase 0.52 accompanied by a rise in the optical veiling (measured between 5400 and 6700~\AA) from 0.2 at phase zero to between 0.4 and 0.7 during the occultation phase.

\begin{figure}
\epsscale{1.2}
\plotone{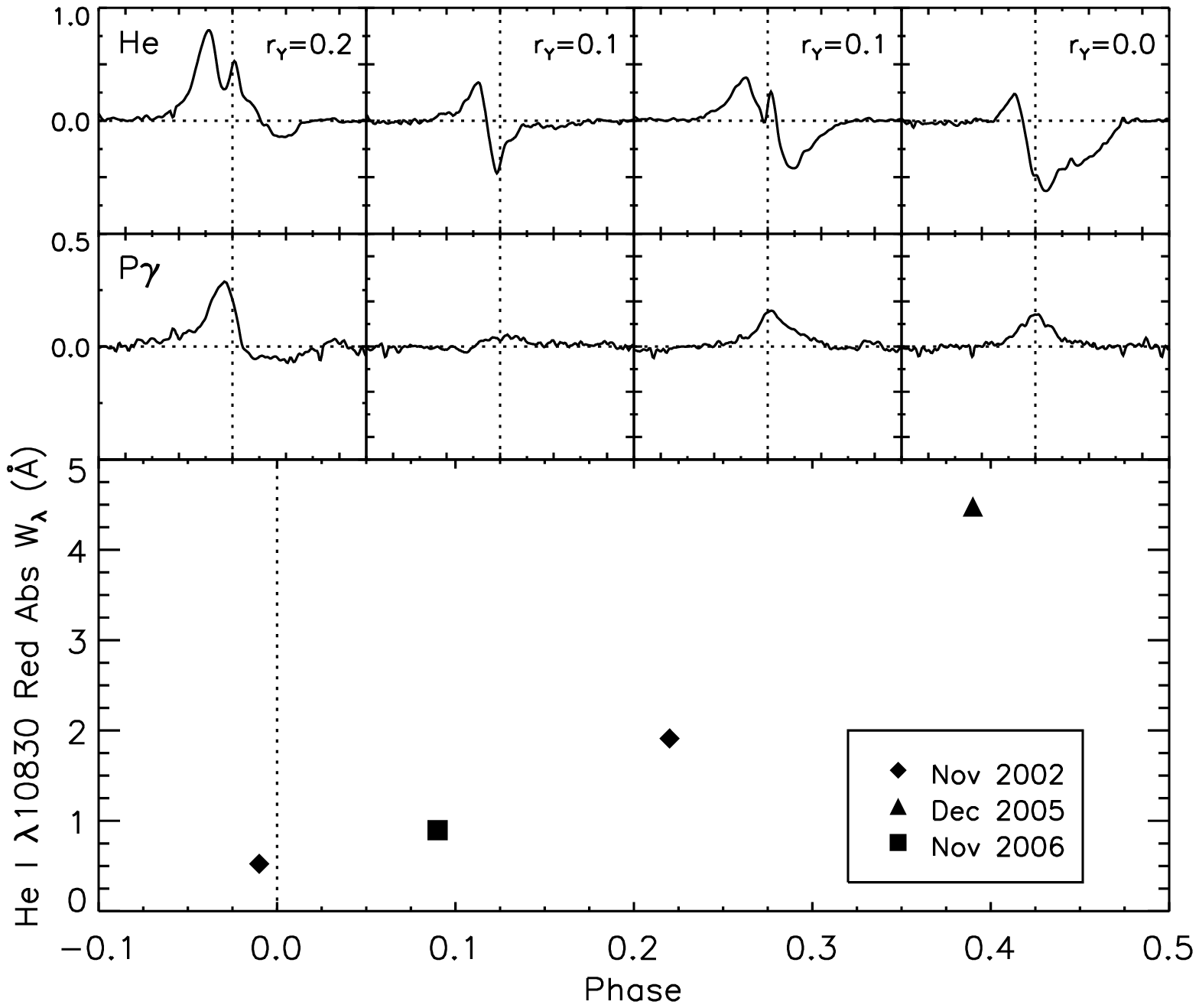}
\figcaption{Relation between red absorption and rotational phase for 4 spectra of AA Tau (phased from \citealt{bou07b}).  The \helium\ and \pgamma\ profiles of AA Tau are shown above the corresponding equivalent width of the red absorption at \helium\ for each phase. Profiles are labeled with the simultaneous 1-$\micron$ veiling, and their velocity axes run from -500 to 500~\kms.  For the helium line, the flux axis runs from -1 to 1, while for \pgamma, it runs from -0.5 to 0.5 to elucidate the morphology of the weaker profiles.\label{f.aavar}}
\epsscale{1}
\end{figure}

To see whether our 1-$\micron$ data of AA Tau are consistent with this picture, we adopt the 8.22-day rotation period, assign phase 0.51 to HJD 2,453,308 \citep{bou07b}, and convert our observation dates to rotation phases. Figure~\ref{f.aavar} shows the \helium\ and \pgamma\ profiles and veilings for each of our 4 observations, corresponding to projected phases from $\sim$~0 to 0.4. The figure also plots the equivalent width of \helium\ red absorption against the derived phase, where each point is roughly aligned with its corresponding profiles.  While \helium\ red absorption appears at all phases, it is weakest near phase zero and increases steadily to phase 0.39. The velocity at the red absorption edge (\vred) also varies, increasing from 250~\kms\ near phase zero to 310~\kms\ at phase 0.39.  In contrast, \pgamma\ shows red absorption only once, close to phase zero, when the 1-$\micron$ veiling is also highest.  If the phasing from the Bouvier epoch is accurate, then the deepest and widest red absorption at \helium\ occurs at the phase associated with maximum accretion effects in the optical when the line of sight pierces the disk warp and the accretion shock. However, this would then mean that the 1-$\micron$ veiling and the \pgamma\ red absorption are out of phase with respect to optical veilings and profiles. Whether or not this phase projection is accurate, the profile sequence for \helium\ and \pgamma\ provides another illustration of the very different kinds of information about the accretion flow that can be inferred from these two lines. Clearly, time-monitoring studies at 1 $\micron$ will be revealing!

\section{SCATTERING MODELS AND COMPARISON TO OBSERVATIONS\label{s.models}}

Radiative transfer models of hydrogen lines arising in the accretion flow \citep {muz01,sym05,kur06,kur08} have been successful in reproducing general characteristics of these lines in some stars.  However, the models assume all of the hydrogen emission arises in the funnel flow, they depend on an assumed temperature in the flow that is not well understood, and they are limited in their ability to constrain the accretion geometry.  In this section, we take a new approach to understanding CTTS accretion flows, {\em modeling the scattering of continuum photons by \helium\ in the infalling gas}.  The lower level of the \helium\ transition (\helower) is 21~eV above the ground state, restricting its formation to regions near the star where the ionizing photon flux is high.  Further, since the only permitted transition downward from the upper level (\heupper) is emission of a $\lambda$10830 photon, we model this line as resonance scattering.

We first lay out the assumptions of our model, in which the accretion geometry is the commonly adopted dipolar flow, a geometrically flat disk is truncated by the innermost field lines, and all accreting field lines terminate in an accretion shock of uniform temperature at the stellar photosphere that generates a continuum excess observable as veiling. We then compare profiles generated from these models to the observed \helium\ red absorption profiles.  The basic dipolar flow is found lacking in a significant number of objects, so we then explore modifications to this geometry that better explain these observations.

\subsection{Basic Dipolar Flow}

We first consider an axisymmetric dipolar field in which the stellar magnetic and rotational axes are aligned and an opaque accretion disk extends from an initial radius $R_i$ to infinity in the equatorial plane.  The outline of the overall structure of accretion from the disk to the star is completely specified by two parameters, although there is some flexibility in which two parameters we choose.  One pair is $R_i$ and $R_f$, which indicate the range in radial distance from the star, i.e., $R_i\le R\le R_f$, over which the dipolar field lines that participate in accretion are distributed over the disk.  Alternatively, we can specify $\theta_i$ and $\theta_f$, which mark the range in polar angle, i.e., $\theta_f\le\theta\le\theta_i$ and $\left(\pi-\theta_i\right)\le\theta\le\left(\pi-\theta_f\right)$, where the same field lines are distributed at the stellar surface.  The relation between the two pairs is apparent from the dipolar field structure, which stipulates that \begin{equation}R_{i,f}=R_*/\sin^2\theta_{i,f}.\label{e.dipole}\end{equation}

A third pair of parameters is $F$ and $R_0$, where $F=\cos\theta_f-\cos\theta_i$ is the fraction of the stellar surface outlined by the above field lines, and $R_0=R_*/\sin^2\theta_0$, with $\cos\theta_f-\cos\theta_0=\cos\theta_0-\cos\theta_i$, marks the approximate median radius of where the accretion flow originates in the disk. The median field line originating at $R_0$ will thus correspond to a median polar angle on the star of $\theta_0$.  We find this pair of parameters to be instructive, since if the whole geometric structure is fully occupied by accreting gas, then $F$ will equal $f$, the filling factor of the accretion shock.  In this subsection, we consider $F=f$ and thus use $f$ to indicate both the fraction of the stellar surface outlined by the overall flow structure and the filling factor of shocked gas at the terminus of accreting field lines, as in previous work by others.

\begin{figure}
\epsscale{1.2}
\plotone{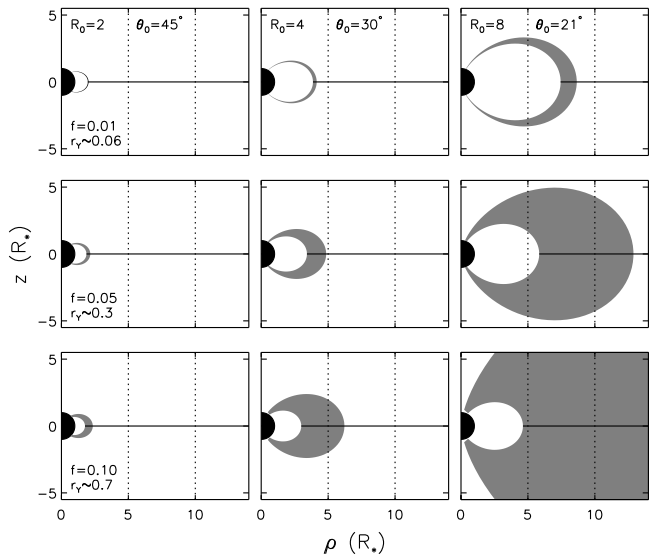}
\figcaption{Schematic representations of accretion geometries used in scattering calculations for dipoles with $F=f$.  The star is black, the accretion flows are gray, and the disk is the solid line in the equatorial plane.  Each column shows a different $R_0$ ($\theta_0$), and each row shows a different $f$ with a corresponding $r_Y$.  We note that in the extreme case of $R_0=8R_*$ and $f=0.1$, the accreting field lines thread the disk out to 35~$R_*$, so the shading in the lower right panel extends far beyond the figure boundary.\label{f.geom}}
\epsscale{1}
\end{figure}

The modeled values of $R_0$ and $f$ are chosen to sample the full range of plausible accretion flow sizes and filling factors. The values of $R_0$, taken to be 2, 4, and 8~$R_*$, are consistent with the understanding that the accretion flow arises near the star-disk corotation radius \citep{gho78,kon91,shu94}.  The values of $f$, taken to be 0.01, 0.05, and 0.1, cover the range found from shock models of CG.   The upper section of Table~\ref{t.space} lists the 9 modeled combinations of $R_0$ ($\theta_0$), $f$ ($=F$), and the associated ranges $\left(R_i, R_f\right)$ over which material leaves the disk.  These configurations are visualized in Figure~\ref{f.geom}.  (The lower half of the table lists cases with $F\ne f$, which will be explored beginning in Section 4.2.)  Three cases correspond closely to geometries used in previous models of magnetospheric accretion \citep{muz01}.  The case with $R_0=4R_*$ and $f=0.01$ approximates their SN (small/narrow) case, the case with $R_0=4R_*$ and $f=0.05$ approximates their SW (small/wide) case, and the case with $R_0=8R_*$ and $f=0.01$ approximates their LW (large/wide) case.

\begin{deluxetable}{ccccccc}
\tablecaption{Model Magnetospheric Geometry Parameters\label{t.space}}
\tablewidth{3in}
\tablecolumns{7}
\tablehead{\colhead{$R_0$} & \colhead{$\theta_0$} & \colhead{$F$} & \colhead{$f$} & \colhead{$r_Y$} & \colhead{$R_i$} & \colhead{$R_f$} \\ \colhead{(1)} & \colhead{(2)} & \colhead{(3)} & \colhead{(4)} & \colhead{(5)} & \colhead{(6)} & \colhead{(7)}}
\startdata
\cutinhead{Undiluted}
2 & 45   & 0.01 & 0.01 & 0.06 & 1.97 & 2.03 \\
  &      & 0.05 & 0.05 & 0.33 & 1.87 & 2.16 \\
  &      & 0.10 & 0.10 & 0.70 & 1.76 & 2.34 \\
4 & 30   & 0.01 & 0.01 & 0.06 & 3.87 & 4.14 \\
  &      & 0.05 & 0.05 & 0.33 & 3.42 & 4.85 \\
  &      & 0.10 & 0.10 & 0.70 & 2.99 & 6.22 \\
8 & 20.7 & 0.01 & 0.01 & 0.06 & 7.44 & 8.65 \\
  &      & 0.05 & 0.05 & 0.33 & 5.84 & 12.9 \\
  &      & 0.10 & 0.10 & 0.70 & 4.63 & 34.5 \\
\cutinhead{Diluted}
2 & 45   & 0.05 & 0.01 & 0.06 & 1.87 & 2.16 \\
  &      & 0.10 & 0.01 & 0.06 & 1.76 & 2.34 \\
  &      & 0.20 & 0.01 & 0.06 & 1.58 & 2.87 \\    
4 & 30   & 0.05 & 0.01 & 0.06 & 3.42 & 4.85 \\
  &      & 0.10 & 0.01 & 0.06 & 2.99 & 6.22 \\
  &      & 0.20 & 0.01 & 0.06 & 2.42 & 15.0 \\
8 & 20.7 & 0.05 & 0.01 & 0.06 & 5.84 & 12.9 \\
  &      & 0.10 & 0.01 & 0.06 & 4.63 & 34.5 \\
\enddata
\tablecomments{Col.~1: Fiducial disk coupling radius ($R_*$); Col.~2: Stellar impact angle in degrees from the pole; Col.~3: Fraction of the star over which the full range of magnetospheric footpoints is distributed; Col.~4: Filling factor of accretion shocks on the stellar surface; Col.~5: Approximate 1-$\micron$ veiling (eq.~[\ref{e.veil}]); Col.~6: Innermost radius at which accreting material leaves the disk, also the disk truncation radius ($R_*$); Col.~7: Outermost radius at which accreting material leaves the disk ($R_*$).  $R_i$ and $R_f$ follow directly from $R_0$ and $F$.}
\end{deluxetable}

For each model the veiling $r_\lambda$ from the associated accretion shock, defined as the ratio of the continuum excess flux $F_v$ to the stellar flux $F_*$, is determined by the blackbody temperatures of the star and hot gas from the shock, the magnitude of $f$, and the viewing angle.  In all cases we assume a $T_*=4000$~K blackbody for the stellar continuum and a $T_v=8000$~K blackbody for the continuum from the accretion shock.  This is a typical value found by CG from continuum excesses shortward of 0.5 $\micron$, although values as high as 10,000 K or as low as 6000 K are sometimes indicated. The veiling at wavelength $\lambda$~is \begin{equation}r_\lambda\equiv\frac{F_v}{F_*}\approx\frac{I_\lambda^{bb}\left(T_v\right)}{I_\lambda^{bb}\left(T_*\right)}\left(\frac{f}{1-f}\right)\label{e.veil},\end{equation} where the approximate equality arises from setting the ratio of the projected areas of the two continua perpendicular to the line of sight, which depends on viewing angle, to simply $f/(1-f)$.  Over the full range of viewing angle, the observed $r_\lambda$ for the same $f$ can vary by a factor of a few (see Section 4.1.1).

The approximate value for $r_Y$ from equation~(\ref{e.veil}), without including the effect of viewing angle, is identified in Figure~\ref{f.geom} for each of the 3 values of $f$.  With our assumed temperatures, the ratio of the blackbody intensities from the veiling continuum and the photosphere $I_\lambda^{bb}\left(T_v\right)/I_\lambda^{bb}\left(T_*\right)$ is 24.5 at $\lambda=5700$~\AA\ and 6.3 at $\lambda=1.08$~\micron, so that for a typical observed $f=0.01$, the approximate veilings at these wavelengths are $r_V=0.25$ and $r_Y=0.06$. The corresponding ratio of $r_V/r_Y\sim4$ is preserved for all $f$ and is independent of viewing angle.  

The velocity of the flow has contributions from both free-fall and rotation.  The free-fall speed at a distance $r$ from the star along a field line threading the disk at $R$ is given by equation~(\ref{e.speed}).  Since the gas follows the field lines, the velocity vector takes the form \begin{equation}{\mathbf v_{ff}} =-v_{ff}\left[\frac{3q^{1/2}(1-q)^{1/2}\hat{\mbox{\boldmath $\rho$}}\pm(2-3q)\hat{\mathbf z}}{(4-3q)^{1/2}}\right]\end {equation}\citep{cal92,har94}.  Here $q=\sin^2\theta$, and $(\hat{\mbox{\boldmath $\rho$}}, \hat{\mbox{\boldmath $\phi$}}, \hat{\mathbf z})$ are unit vectors in the cylindrical coordinate system.  Above the equatorial plane, the plus sign applies, while below the equatorial plane, the minus sign applies so that the flow is always from the disk to the star.

For the rotational component of the flow, the magnetosphere is assumed to rotate rigidly with velocity \begin{equation}{\mathbf v_{\phi}}=v_*\frac{\rho}{R_*}\hat{\mbox{\boldmath $\phi$}},\end{equation} where $v_*$ is the rotation speed of the star at its equator, assumed here to be $0.05~V_{\rm esc}$, or 15~\kms\ when $V_{\rm esc}=300$~\kms, a typical value for TTS \citep[and references therein]{reb04}, and $\rho$ is the cylindrical radial distance of a point from the rotation axis.  Since the rotational motion is for the most part transverse to the line of sight for the absorbing gas seen projected in front of the star, it has a very small effect on the absorption part of the line profile.

The flow scatters continuum photons, which arise from the star and the accretion-heated photosphere.  To maximize the red absorption, the $\lambda$10830 transition in the accreting flow is assumed to be optically thick.  A rectangular line absorption profile with a 10~\kms\ half-width is adopted to account for thermal and turbulent broadening.  Thus, if a particular ray from a point on the stellar surface intersects the accreting flow such that the projection of the gas velocity along the ray extends from $v_{\rm min}$ to $v_{\rm max}$, then continuum photons from ($v_{\rm min}-10$~\kms) to ($v_{\rm max}+10$~\kms) are scattered.  Because spontaneous emission is the dominant de-excitation route of the $\lambda$10830 upper state (\heupper) in comparison with other decay, collisional, or ionization processes, the photon absorption and subsequent re-emission is in effect a resonant scattering process if the small fine-structure energy differences among the three sub-levels are ignored. Rather than following the photon path in detail (e.g., with a Monte Carlo simulation), we simply assume a single scattering in which the absorbed photon is re-emitted isotropically with the appropriate Doppler shift, and it either hits the star; hits the opaque, flat disk; or escapes the system.  While this is inconsistent with the assumption of an opaque line, we find that the exact contribution to the observed profile from the scattered photons has no significant bearing on our conclusions (see subsequent sections), so the extra effort is unwarranted.

The emergent spectrum at a particular viewing angle $i$ is made up of photons that, either because they are not absorbed or they are scattered, escape into a solid angle bin centered on $i$.  For a random selection of $i$ over $4\pi$ steradians, $\cos i$ is uniformly distributed.  Considering five viewing angles, we choose $\cos i=0.9$, 0.7, 0.5, 0.3, and 0.1, or $i=26^\circ$, $46^\circ$, $60^\circ$, $73^\circ$, and $84^\circ$.

\subsubsection{Basic Dipolar Flow: Results}

\begin{figure}
\epsscale{1.2}
\plotone{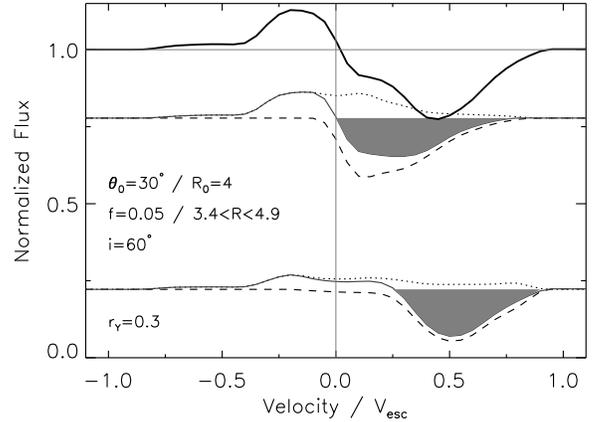}
\figcaption{Example scattering profile for dipolar infall with $\theta_0=30^\circ$ and $F=f=0.05$ (equivalently, $R_0=4R_*$ and $3.4<R/R_*<4.9$), viewed from an angle $i=60^\circ$.  The emergent profile (black line) is the sum of two components: the profile due to photons from the 4000-K stellar continuum (upper gray line) and the profile due to photons from the 8000-K veiling continuum (lower gray line).  The veiling, shown in the lower left, is the ratio of the veiling continuum height to the stellar continuum height.  Each component is further made up of two subprofiles: the absorption profile of the continuum (dashed) and the emission profile of scattered photons that escape toward the line of sight (dotted).\label{f.example}}
\epsscale{1}
\end{figure}

Figure~\ref{f.example} shows an example of the components contributing to the emergent model profile in the case of $R_0=4R_*$ ($\theta_0=30^\circ)$, and $f=0.05$ ($3.4\le R/R_*\le4.9$), viewed at $i=60^\circ$.  The final emergent spectrum, shown in black, is the sum of contributions from the stellar and the veiling (accretion shock) continua, each shown separately in solid gray.  The stellar contribution is the one with a normalized continuum level of 0.78, due to scattering of the 4000-K continuum, while the veiling contribution is the one with a normalized continuum level of 0.22, due to scattering of the 8000-K continuum.  The ratio of these two continua, $0.22/0.78=0.3$, is the 1-$\micron$ veiling $r_Y$, also noted in the figure.  Each solid gray component is further the sum of two subcomponents.  One, shown with a dashed line in each case, is the absorption profile of the respective continuum.  The other, shown with a dotted line in each case, is the emission profile, produced by scattered photons that escape toward the specified line of sight.  The emission subcomponent is both broad and weak, since scattered photons can be red- or blueshifted and because for each photon absorbed, the ensuing emitted photon may hit the disk or star and not escape.  Thus the filling-in of the red absorption by its own associated scattered emission is generally slight.  

Figure~\ref{f.example} also illustrates an important aspect of the models, that the redshifted absorption in the emergent profile is affected differently by scattering of the stellar and the veiling continua.  In this model, with $\theta_0=30^\circ$ and $i=60^\circ$, the line of sight toward the veiling continuum intersects the portion of the accretion flow close to the star where the gas velocity is high (see Fig.~\ref{f.geom}), and scattering of the veiling continuum produces a red absorption that extends from 0.27 to 0.87~\vesc.  In contrast, the line of sight toward the stellar continuum intercepts portions of the accretion column with smaller infall speeds and a smaller velocity component is projected onto the line of sight.  The red absorption thus produced ranges from $\sim 0$ to 0.74~\vesc\ and is also shallower than the one from the veiling continuum.  The resultant absorption profile is thus complex in shape and broader than either of the two individually, and in this case it has a maximum depth of about 20\% into the summed continuum. 

\begin{figure*}
\includegraphics[angle=90,width=\textwidth]{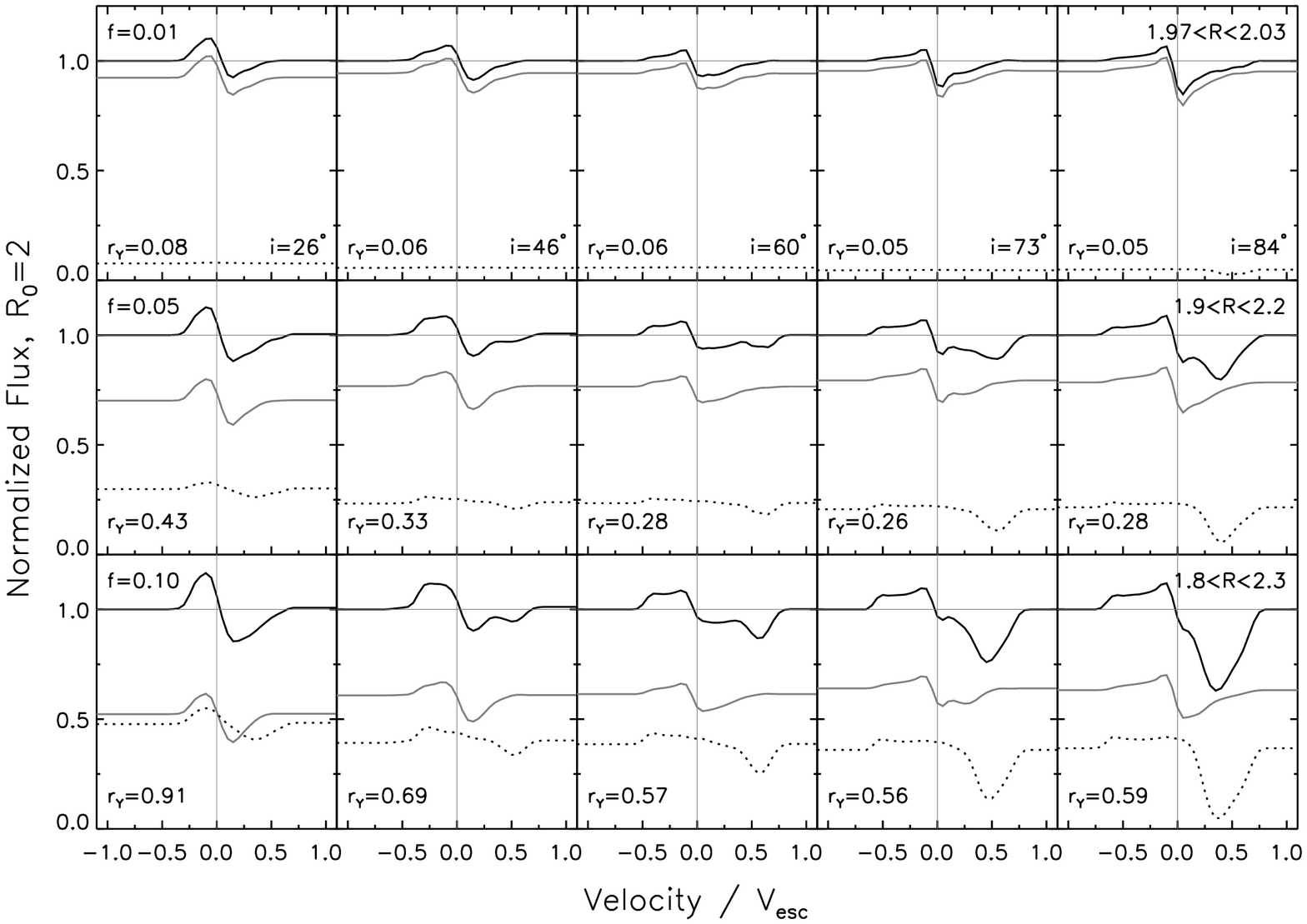}
\figcaption{Scattering profiles for dipolar infall with $R_0=2R_*$ ($\theta_0=45^\circ$) and $F=f$.  
Each row shows a different value of $f$.  Within each row, each panel shows the profile for a different viewing angle and the corresponding 1-$\micron$ veiling $r_Y$.  Emergent profiles (upper black lines) are the sum of the profiles from the stellar continuum ($T=4000$~K; gray lines) and the veiling continuum ($T=8000$~K; dotted lines).  The optical veiling (at $\lambda=5700$~\AA) is approximately 4 times greater than $r_Y$.\label{f.simr2}}
\end{figure*}

\begin{figure*}
\includegraphics[angle=90,width=\textwidth]{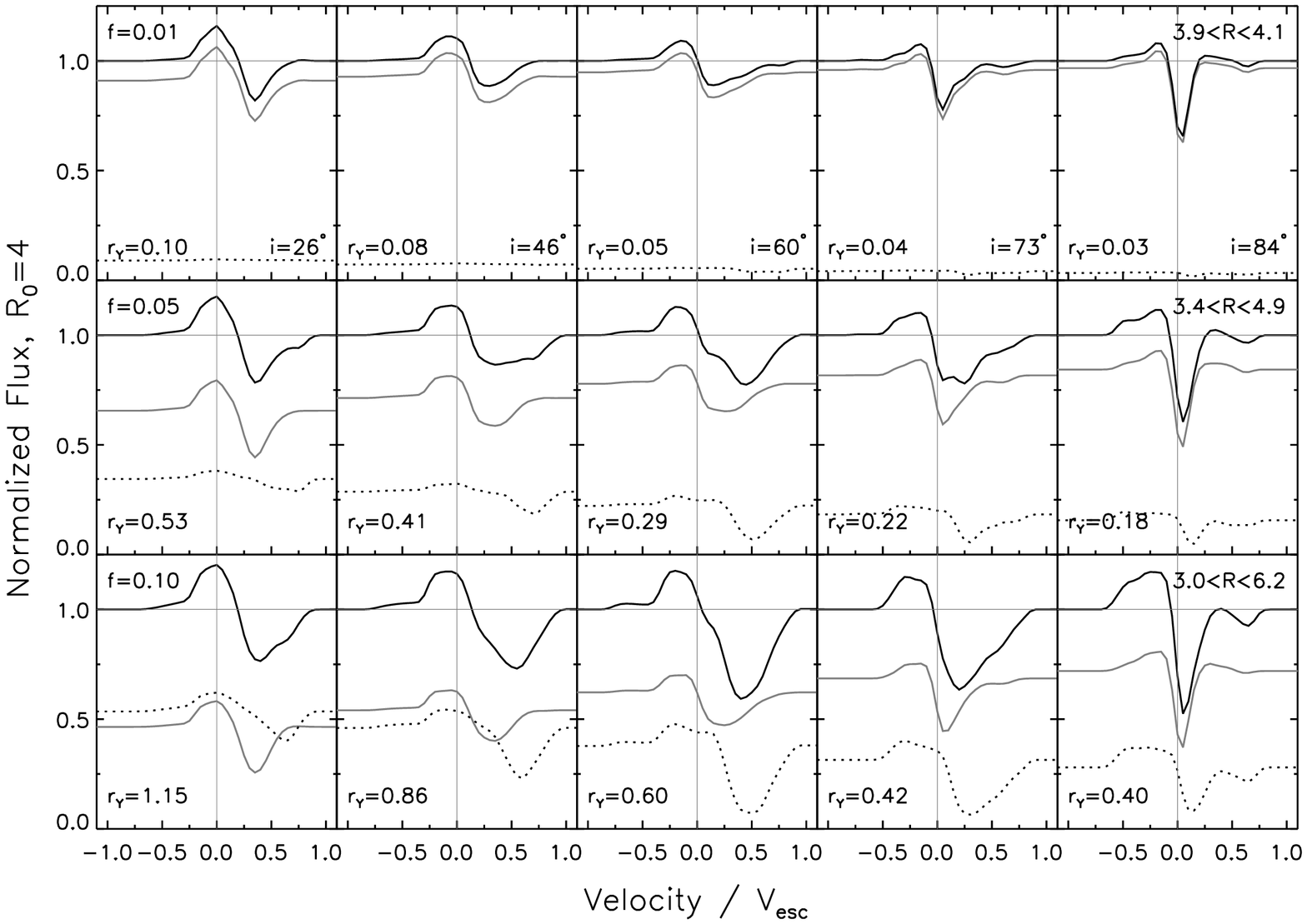}
\figcaption{Scattering profiles for dipolar infall with $R_0=4R_*$ ($\theta_0=30^\circ$) and $F=f$, as in Figure~\ref{f.simr2}. \label{f.simr4}}
\end{figure*}

\begin{figure*}
\includegraphics[angle=90,width=\textwidth]{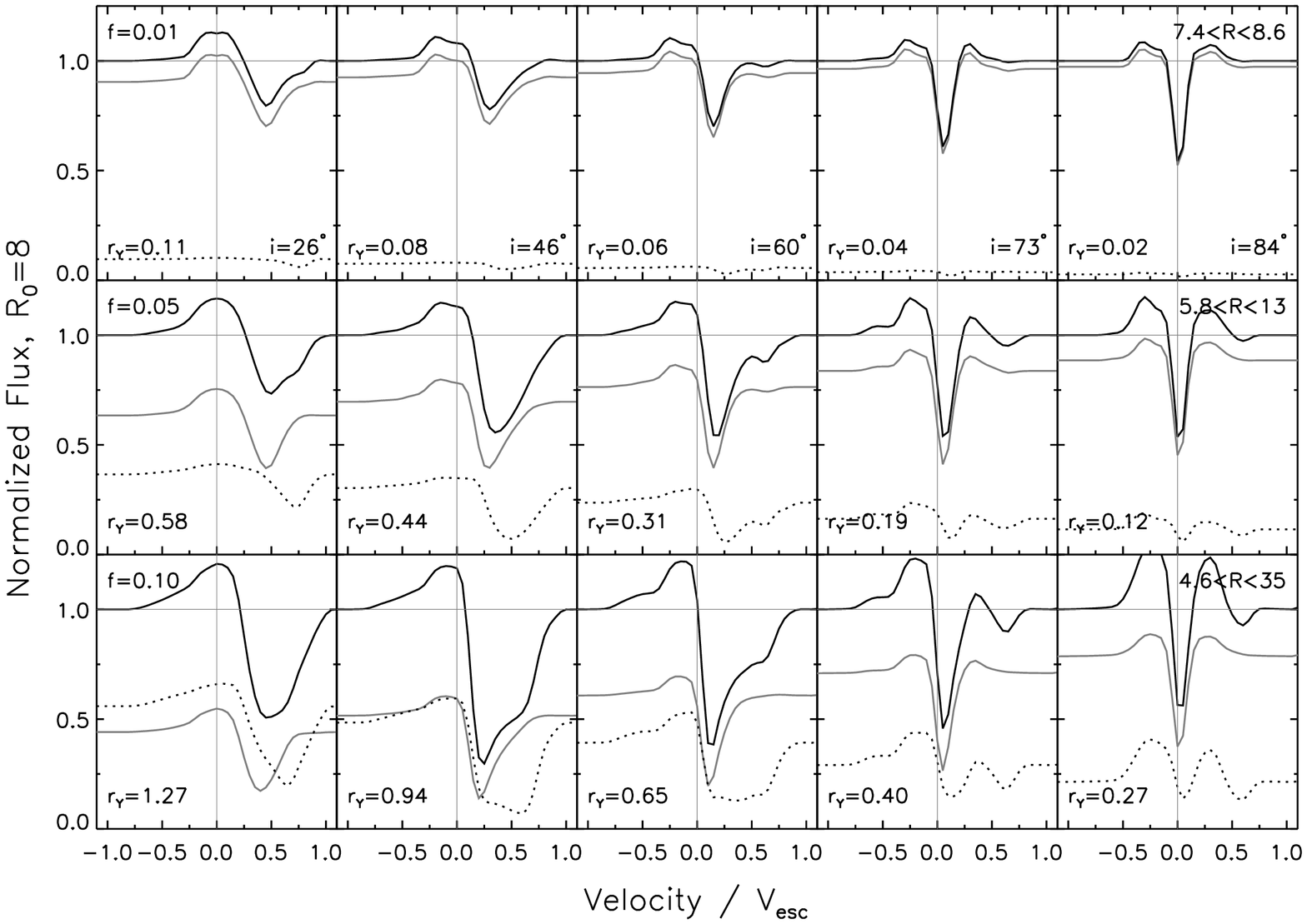}
\figcaption{Scattering profiles for dipolar infall with $R_0=8R_*$ ($\theta_0=20.7^\circ$) and $F=f$,  as in Figure~\ref{f.simr2}.\label{f.simr8}}
\end{figure*}

Figures~\ref{f.simr2}, \ref{f.simr4}, and \ref{f.simr8} show the full range of model profiles for the three chosen values of $R_0$: 2, 4, and 8~$R_*$ respectively (corresponding to $\theta_0=45^\circ$, $30^\circ$, and $20.7^\circ$).  In each figure, the three rows show, from top to bottom, the three selections of $f=0.01$, 0.05, and 0.1.  In each row, the five panels show the profiles for the five viewing angles, from $i=26-84^\circ$.  Within each of the 15 panels representing a unique combination of $f$ and $i$, the final emergent profile is shown (solid black curve) along with the separate contributions from scattering of the veiling continuum (dotted curve) and the stellar continuum (solid gray curve), but not the subcomponents of absorption and scattered emission from each continuum source.

We emphasize the following points from these three figures:

1. When $f=0.01$, the red absorption is dominated by scattering of the stellar continuum, generally showing small absorption equivalent widths and velocity widths. The profiles have a strong dependence on inclination, with shallow absorption at small inclinations and narrow, deeper, low-velocity absorption at high inclinations.  

2. The magnitude of the red absorption, measured by either the equivalent width or the maximum depth of absorption, is  sensitive to the parameter $f$.  As $f$ increases, there is both an increase in the veiling continuum and an increase in the coverage of accreting field lines projected in front of the stellar surface for a given $R_0$, enabling the line of sight to each point on the star to intersect more accreting field lines and hence yield a broader range in the projected velocity of the infalling gas.

3. For a given $f$, the red absorption is generally stronger at a larger $R_0$ (smaller $\theta_0$), since the accreting field lines then cover a greater range of solid angles, and the larger span between $R_i$ and $R_f$ produces a broader range in the gas velocity.  However, inclination also plays a role, so that for a given $f$ and $R_0$, the strongest absorption occurs at a line of sight $i$ that parallels the final part of the trajectory of the accretion flow.  From the schematic in Figure~\ref{f.geom}, it can be seen that for $\theta_0=45^\circ$, $30^\circ$, and $20.7^\circ$, the corresponding viewing angle to maximize the red absorption is roughly $i\approx84^\circ$, $60^\circ$, and $46^\circ$ respectively.  Thus an increased $\theta_0$ (smaller $R_0$) requires a higher $i$ for a strong red absorption.  This occurs because the contribution to the absorption from scattering of the veiling continuum is broadest when viewed in a direction parallel to the flow just before it impacts the star. 

4. The observed emission, i.e., the part of the profile above the continuum, is usually weaker than the absorption and is mostly blueshifted.  Only in the extreme but unrealistic case when $R_0=8R_*$ ($\theta_0=20.7^\circ$) with high $f$ (0.05 or 0.1) and excessive $R_f$ (13 and 35 $R_*$ for $f=0.05$ and 0.1 respectively) does a double-peaked profile result when viewed close to edge on. The resulting accretion flow has a large solid angle, and since it is assumed to be in corotation with the star, the rotational broadening is considerable.  This situation was included to complete our chosen parameter space and is not realistic.

5.  The veiling ($r_\lambda$) from the 8000-K accretion zone depends on $f$, $\theta_0$ and $i$.  The dependence on $f$ is obvious, since $r_\lambda$ scales almost linearly with $f$ (eq.~[\ref{e.veil}]).  The dependences on $\theta_0$ and $i$ arise through their influences on the projected area of each continuum source.  The parameter $\theta_0$ signifies the orientation of the veiling continuum, hence its direct effect on the veiled area viewed.  Although less sensitive to the viewing angle, the projected stellar continuum area changes because of the presence of the disk extending from $R_i$ (dependent on $\theta_0$ and $f$) to infinity.  For both $\theta_0=20.7^\circ$ and $30^\circ$, $r_Y$ at a given $f$ drops monotonically as $i$ increases from pole-on to edge-on, by a factor of 5 and 3 respectively.  At $\theta_0=45^\circ$, $r_Y$ varies less, dropping by a factor of 1.6 from $i=26^\circ$ to $73^\circ$, then increasing slightly toward $i=84^\circ$.  For example, when $R_0=4R_*$ ($\theta_0=30^\circ$) and $f=0.1$, $r_Y$ ranges from 0.4 to 1.15 and $r_V$ ranges from 1.5 to 4.5 with viewing angle.  

In sum, a strong red absorption extending to high velocities, like those observed, requires $f$ (and thus $r_\lambda$) to be large so the contribution from scattering of the veiling continuum is enhanced and the angular extent of the flow on the star is increased.  Strong, broad absorption is also more likely when $R_0$ is large and the line of sight parallels the accretion flow close to the star. The relation between absorption magnitude and $r_\lambda$ is a crucial test of the dipolar accretion model, as we will show in the following subsection when we compare the observations to our model profiles.\\

\subsubsection{Basic Dipolar Flow: Comparison to Observations}

In comparing our models with observed profiles, we focus on the red absorptions, since the small emission at blueward velocities expected from scattering in the funnel flow will often be overwhelmed by additional sources of emission, such as scattering and in-situ emission from a wind. The red absorptions are evaluated in context with the observed veiling, which is the basis for evaluating $f$. It is immediately apparent that there is a mismatch between the model profiles in Figures~\ref{f.simr2}, \ref{f.simr4}, and \ref{f.simr8} and the observed spectra in Figure~\ref{f.redabs}, since the majority of CTTS are known to have $f \lesssim0.01$ (CG) while model sequences for $f=0.01$ (implied $r_Y \sim 0.06$ and $r_V \sim 0.25$) have shallow and/or narrow red absorptions bearing little resemblance to the ensemble of broad and deep observed \helium\ profiles.

\begin{figure}
\epsscale{1.2}
\includegraphics[angle=90,width=0.5\textwidth]{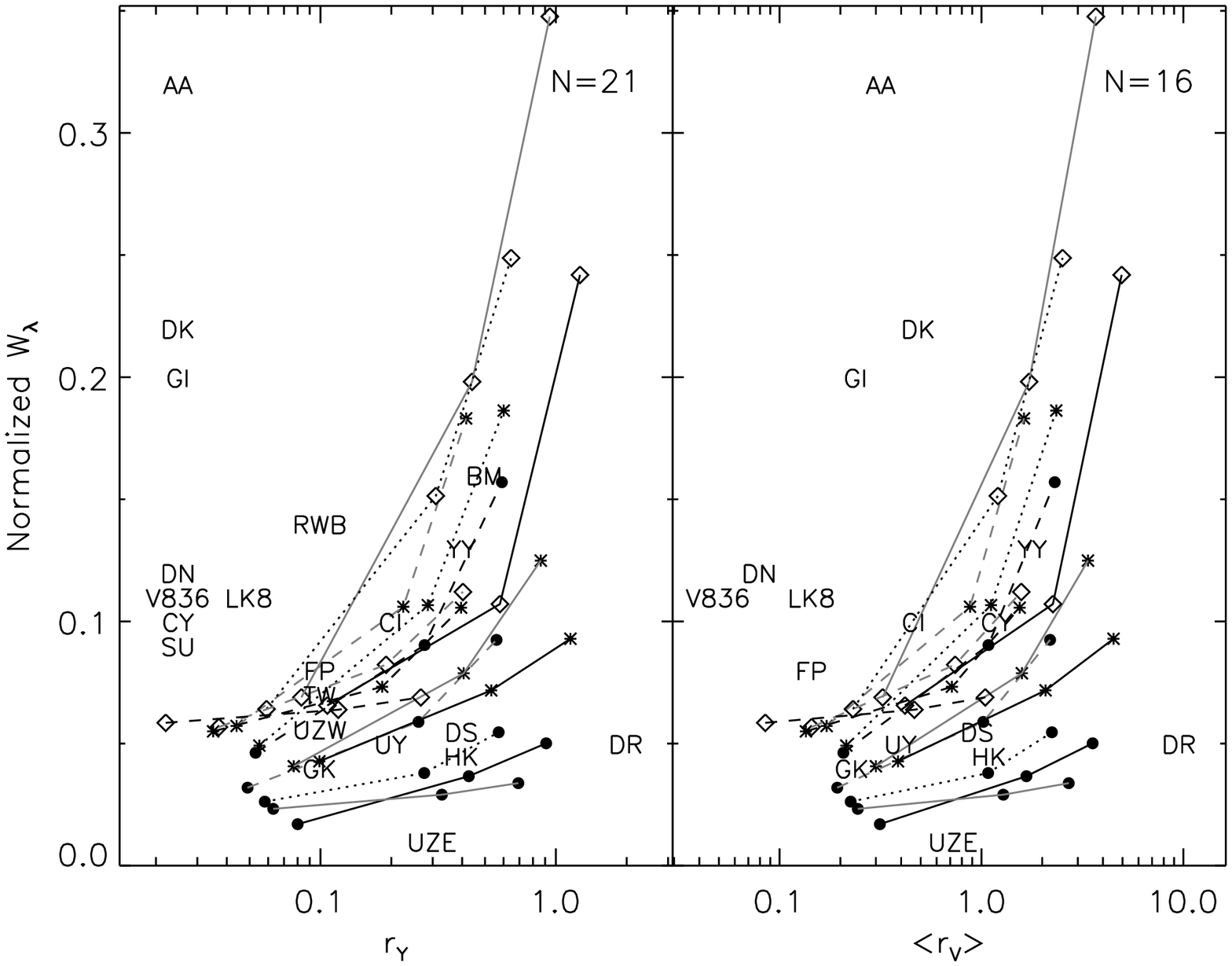}
\figcaption{Comparison of the red absorption equivalent width (normalized to the escape velocity) to the 1-$\micron$ veiling (left) and the non-simultaneous average optical veiling (right; from HEG) for basic dipolar models and the profiles from the reference sample.  The model properties appear as lines connected by symbols.  Each symbol type is for a different $R_0$ / $\theta_0$, with circles for $R_0=2R_*$, asterisks for $R_0=4R_*$, and diamonds for $R_0=8R_*$.  Each line type is for a different viewing angle, with solid black for $26^\circ$, solid gray for $46^\circ$, dotted for $60^\circ$, dashed gray for $73^\circ$, and dashed black for $84^\circ$.  Along a line, symbols indicate filling factors $f=0.01$, 0.05, and 0.10, always increasing toward increasing veiling.  Since the veiling axes are logarithmic, stars with no detected 1-$\micron$ veiling are placed at $r_Y=0.025$.\label{f.comp}}
\epsscale{1}
\end{figure}

A more explicit demonstration of the limitation of the models can be made from a quantitative comparison between the equivalent width and veiling for model and observed profiles. This comparison requires normalizing the observed profiles to their respective escape velocities, since in the models all velocities are in units of the escape velocity. The normalized equivalent width, $W_\lambda'=W_\lambda/V_{\rm esc}$, has an intuitive interpretation: it is simply the fraction of the continuum absorbed between rest and the escape velocity, with a value of 1 indicating total absorption over the entire range. Figure~\ref{f.comp} compares the normalized red absorption equivalent width $W_\lambda'$ to $r_Y$ and $r_V$ for both models and observations of the reference sample.  In the models we have assumed that the excesses at both $Y$ and $V$ arise from an accretion shock that emits an 8000-K blackbody continuum.  This is known to be a valid assumption for optical veilings, and while the $r_V$ data points are not simultaneous with the observed absorption profiles, the fact that all but one of the objects with $r_Y=0$ also have low $r_V$ indicates that this is a reasonable approach.  Unless the $r_V$ values for these objects were all a factor of 5 to 10 higher when the \helium\ profiles were obtained than when the HEG data were obtained, the two panels together clearly indicate that only a fraction of the observed data lie within the realm of the model results: those with weak red absorption and small veiling or those with modest red absorption and intermediate veiling. There is a glaring discrepancy between models and observations for stars with large $W_\lambda'$ and small $r_Y$. 

\begin{figure}
\epsscale{1.2}
\plotone{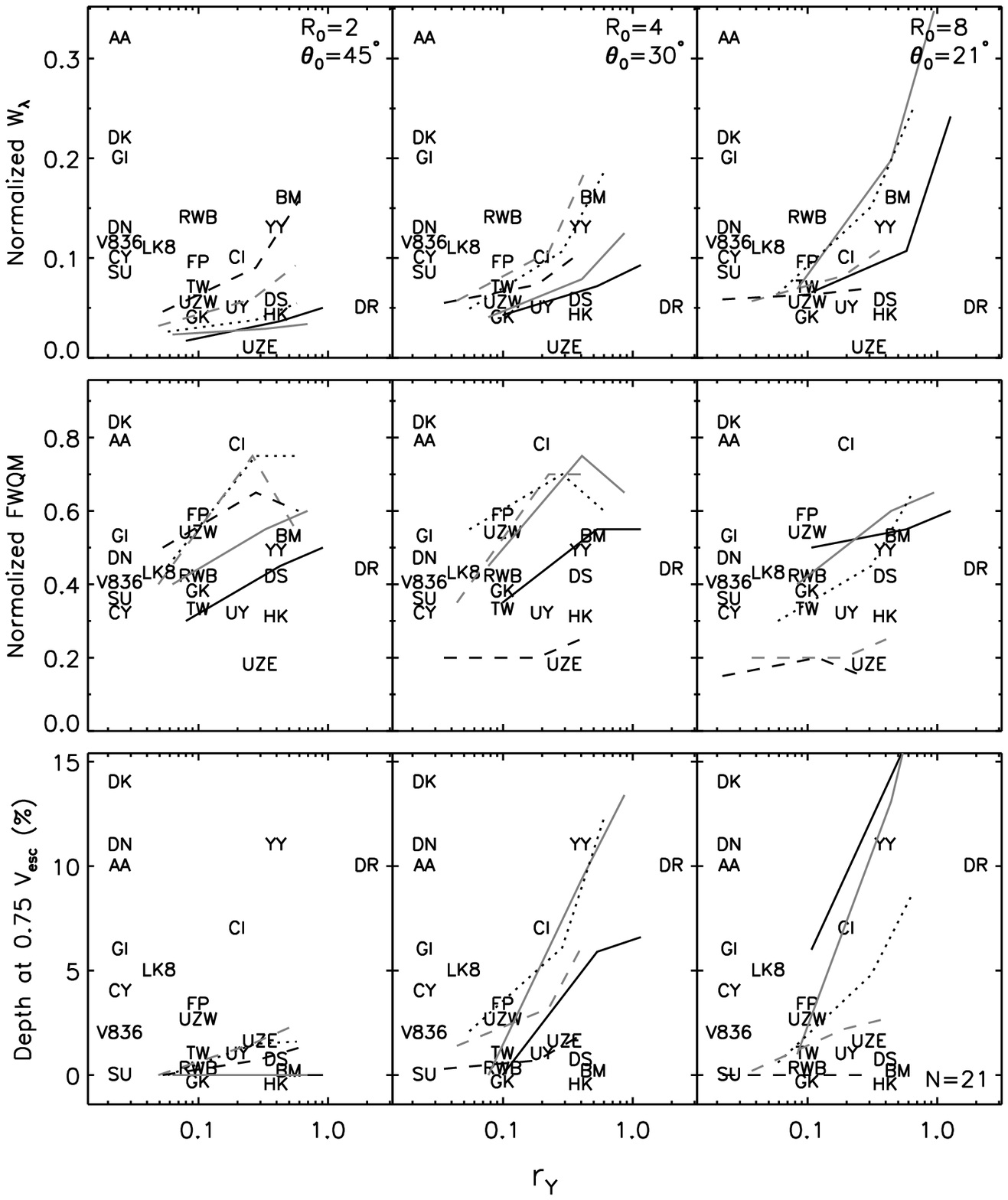}
\figcaption{Comparison of 3 properties of the red absorption to the 1-$\micron$ veiling for basic dipolar models and the 21 profiles from the reference sample.  In each column, the observed parameters are the same, but model parameters appear only for the indicated $R_0$ / $\theta_0$ combination.  The model parameters for the three filling factors $f=0.01$, 0.05, and 0.10 at a particular viewing angle are connected by lines, and the correspondence between line type and viewing angle is the same as in Figure~\ref{f.comp}.  As before, stars with no detected 1-$\micron$ veiling are placed at $r_Y=0.025$.\label{f.comp2}}
\epsscale{1}
\end{figure}

The values for $W_\lambda'$ and adopted escape velocities are listed in Table~\ref{t.normal} along with two additional properties of the red absorption: the normalized width and the depth.   The normalized full-width at quarter-minimum, FWQM$'$, is the width measured at one quarter of the absorption minimum as a fraction of the escape velocity.  The depth of the absorption component at 0.75~\vesc\ normalized to 100\%, $D_{0.75}$, was chosen since it is sensitive to the infall geometry close to the star. As with $W_\lambda'$, a number of stars indicate a discrepancy with the basic dipole model, where objects with small veiling frequently have both FWQM$'$ and $D_{0.75}$ much larger than can be accounted for with the models. This is illustrated in Figure~\ref{f.comp2}, which shows the comparison of the velocity-normalized equivalent width, velocity-normalized line width, and high-velocity depth to 1-$\micron$ veiling for observations and models.

\begin{deluxetable}{lcccc}
\tablecaption{Measurements of Red Absorption in Velocity-Normalized \helium\ Profiles\label{t.normal}}
\tablewidth{3in}
\tablehead{\colhead{Object} & \colhead{$V_{\rm esc^*}$\tablenotemark{a}} & \colhead{$W_\lambda'$} & \colhead{FWQM$'$} & \colhead{$D_{0.75}$} \\ \colhead{(1)} & \colhead{(2)} & \colhead{(3)} & \colhead{(4)} & \colhead{(5)}}
\startdata
AA Tau\dotfill   & 390   & 0.32 & 0.79 & 10 \\
BM And\dotfill   & 510   & 0.16 & 0.53 &  0 \\
CI Tau\dotfill   & 370   & 0.10 & 0.78 &  7 \\
CY Tau\dotfill   & 310   & 0.10 & 0.32 &  4 \\
DK Tau\dotfill   & [380] & 0.22 & 0.84 & 14 \\
DN Tau\dotfill   & 300   & 0.11 & 0.47 & 11 \\
DR Tau\dotfill   & [360] & 0.05 & 0.44 & 10 \\
DS Tau\dotfill   & 570   & 0.05 & 0.42 &  0 \\
FP Tau\dotfill   & 200   & 0.08 & 0.60 &  3 \\
GI Tau\dotfill   & 450   & 0.20 & 0.53 &  6 \\
GK Tau\dotfill   & 350   & 0.04 & 0.40 &  0 \\ 
HK Tau\dotfill   & 320   & 0.05 & 0.31 &  0 \\
LkCa 8\dotfill   & 370   & 0.11 & 0.43 &  5 \\
RW Aur B\dotfill & 580   & 0.14 & 0.40 &  0 \\
SU Aur\dotfill   & 490   & 0.09 & 0.37 &  0 \\
TW Hya\dotfill   & 520   & 0.07 & 0.33 &  1 \\
UY Aur\dotfill   & 400   & 0.05 & 0.32 &  1 \\
UZ Tau E\dotfill & 340   & 0.01 & 0.18 &  1 \\
UZ Tau W\dotfill & 260   & 0.07 & 0.54 &  3 \\
V836 Tau\dotfill & 440   & 0.11 & 0.39 &  2 \\
YY Ori\dotfill   & [430] & 0.13 & 0.49 & 11 \\
\enddata
\tablecomments{For the reference sample only.  Col.~2: Adopted escape velocity (\kms); Col.~3: Equivalent width of velocity-normalized absorption (dimensionless); Col.~4: Full-width at quarter-minimum of velocity-normalized absorption (dimensionless); Col.~5: Depth at 75\% of the escape velocity as a percentage of the continuum.} 
\tablenotetext{a}{Brackets indicate $V_{\rm red}/V_{\rm esc}>1$, so we assume $V_{\rm esc^*}=V_{\rm red}/0.9$.}
\end{deluxetable}

This comparison demonstrates that a fraction of CTTS with subcontinuum red absorption at \helium\ have red absorptions too strong to be accounted for by magnetospheric  accretion in a basic dipole, where the filling factor of the flow on the stellar surface $F$ is equivalent to the filling factor of shocked gas at the terminus of accreting field lines $f$.  The observations that present the greatest challenge to the model are those in which the red absorption is strong ($W_\lambda'\ge0.1$) but the veiling is weak ($r_Y\le0.1$).  This conclusion is robust, since the models have been constructed to produce maximal red absorption for a given $R_0$ and $f$, in that the $\lambda$10830 transition is assumed to be optically thick and the thermal/turbulent broadening has a generous 10~\kms\ half-width.

This conclusion is not compromised by the choice of 8000~K for the temperature of the shock-heated photosphere. This value corresponds to the low end of the temperature range derived from modeling the SEDs of observed continuum excesses \citep{har91,gul98,joh00}.  If higher temperatures were assumed, the veiling for a given $f$ would be even larger, worsening the agreement between the models and the observations.  If we adopted the lowest temperature allowed by the SED models of the optical continuum excess, $T\sim6000$~K, the associated veiling for a given $f$ would be reduced by $\sim2$ at $Y$ and $\sim3$ at $V$.  Figure~\ref{f.comp} demonstrates that shifting all the model results to the left by a factor of 2 in $r_Y$ or 3 in $r_V$ is still insufficient to account for the strong absorptions and low veilings.  We thus conclude that those profiles with strong absorptions and small veilings lie outside the realm of model results for self-consistent dipole flows.

\subsection{Dilution}

A simple way to keep the veiling small and yet have the accretion flow project a broad velocity range  in front of the star is to let the flow arise over a large range of $R$ (thus impacting the star over a large range of $\theta$) but to fill the whole enclosed volume only {\em dilutely} with accreting gas. We now distinguish between $F$, the fractional surface area on the star over which the magnetospheric footpoints are distributed, and $f$, the fractional surface area on the star occupied by accretion shocks at the base of field lines that carry accreting gas.  We define $f'\equiv f/F$ as the fraction of $F$ occupied by all the accretion shocks.  With enough dilution, i.e., $f'$ sufficiently small, $F$ can be large enough to provide the areal coverage over a large velocity range that is necessary for a broad and deep red absorption, while $f=Ff'$ can remain small, as required to produce a low veiling.  One way to achieve this is to postulate a large number of narrow accretion streamlets spatially separated from one another that together impact only a fraction $f'$ of the outlined area $F$.  (We assume the many accretion shocks are dispersed randomly throughout $F$.)  Then, with an intrinsic thermal or turbulent line broadening of $\sim10$~\kms\ associated with each streamlet, photons from the star can intersect a sufficient number of streamlets such that the continuum (stellar or veiling) will be absorbed over the full velocity range specified by the parameter $F$ as though the whole volume were filled.

The concept of many accretion streamlets dilutely filling a volume has the additional advantage of offering a credible explanation for how the lower level of $\lambda$10830 (\helower) is populated over all streamlines. With the difficulty of maintaining a temperature high enough ($\gtrsim2\times10^4$~K) for collisional excitation to the \helower\ level in a freely falling gas, it is likely that photoionization is the excitation mechanism.  Then, if the source of ionizing photons is the accretion shock itself, the much smaller shocked area of an individual streamlet within a diluted flow, as compared to the shocked area of a single undiluted flow, will enable more ionizing photons to escape from the sides and ionize the gas in other streamlets, even at positions far from the star.  Or, if the dominant source of ionizing radiation is located away from the streamlets (e.g., the stellar corona), these photons will be able to penetrate into the volume and ionize individual streamlets as opposed to ionizing just the skin of a single completely filled accretion flow.  Thus, many narrow streamlets dilutely filling a large volume not only yield a deep red absorption from the large coverage area over a broad velocity range of infalling gas, but they also readily account for the ionization of gas at each location in the flow to produce an optically thick $\lambda$10830 transition over the whole velocity range.

\subsubsection{Profiles for Wide, Dilutely Filled Flows}

\begin{figure}
\epsscale{1.2}
\plotone{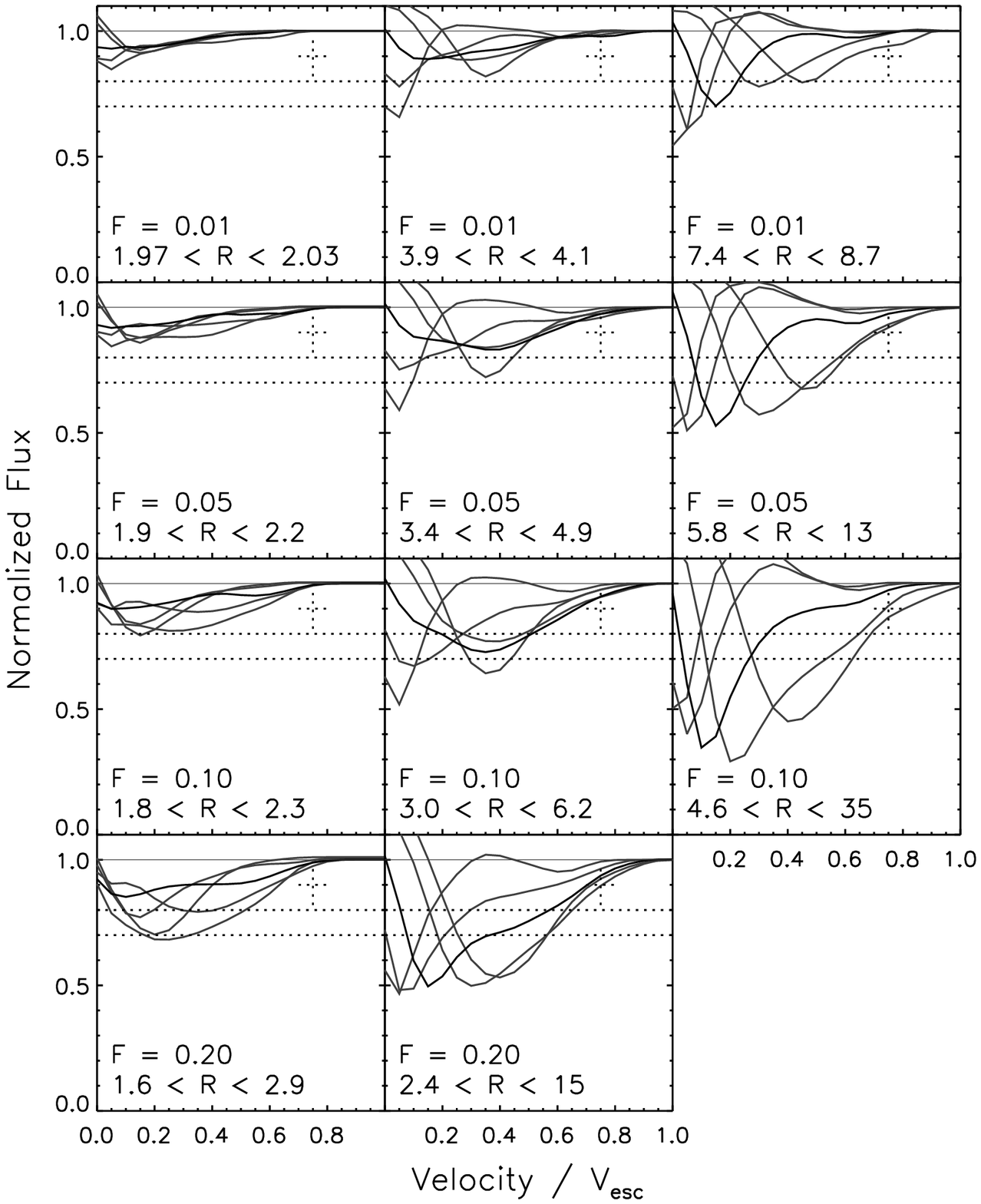} 
\figcaption{Red side of scattering profiles for a series of ``diluted'' dipoles all with $f=0.01$ ($r_Y \le 0.11$) but with $F$ ranging from 0.01 (top row) to 0.2  (bottom row).  From left to right, columns correspond to $R_0=2$, 4, and 8~$R_*$, and the range of $R$ for each $F$ is specified.  The profile sequences for each panel correspond to viewing angles of $26^\circ$, $46^\circ$, $60^\circ$ (black), $73^\circ$, and $84^\circ$.  The top row with $F=f=0.01$ corresponds to the models shown in the upper rows of Figures~\ref{f.simr2}, \ref{f.simr4}, and \ref{f.simr8}.  For comparison with observations, dotted horizontal lines mark depths of 20\% and 30\% and crosses mark a depth at $V/V_{\rm esc}=0.75$ ($D_{0.75}$) of 10\%. \label{f.dilute}}
\epsscale{1}
\end{figure}

We compute scattering profiles for diluted dipole flows for the same 3 geometries shown earlier, with ($R_0$, $\theta_0$) pairs of (2 $R_*$, $45^\circ$),  (4 $R_*$, $30^\circ$), and (8 $R_*$, $20.7^\circ$). We introduce a wider range in $F$, from 0.01 to 0.2, although now all models have $f=0.01$, corresponding to a range in $f'$ from 1 to 0.05. The resulting profiles are shown in Figure~\ref{f.dilute}, and the model parameters are listed in the lower portion of Table~\ref{t.space}.  In the figure, the 3 columns correspond to the 3 $R_0$ values and each row is a common value of $f'$. The degree of dilution increases downward in the figure, with the case for no dilution shown in the top row ($f'=1$ and $f=F$) repeated from Figures~\ref{f.simr2}, \ref{f.simr4}, and \ref{f.simr8}. In subsequent rows the dilution grows as $F$ increases to 0.05, 0.1, and finally 0.2. Each panel shows the superposed profiles for all 5 viewing angles for each $R_0$, $f'$ (or $F$) combination and the corresponding inner and outer radii of the accreting volume, $R_i<R<R_f$. Since the effect of the viewing angle on the profile morphology is roughly independent of dilution, the individual viewing angles can be identified by referring to the earlier figures. We highlight the $i=60^\circ$ profile with a darker line, since this is the most probable viewing angle.

Although all profiles in Figure~\ref{f.dilute} are for $f=0.01$, the associated veilings differ slightly because the accretion-heated area is distributed differently for different values of $F$ and $\theta_0$, leading to slightly different projected areas.  Nonetheless, in all cases, $r_Y<0.11$.  For these small veilings, the absorption, while quite strong when there is significant dilution, is almost entirely due to scattering of the stellar continuum, in contrast to the undiluted models where large veilings and scattering of the veiling continuum were necessary to produce strong absorption.  As dilution increases for a given $R_0$, the red absorptions become increasingly strong and broad, due to the increased areal coverage over a broader range of velocities as the interval between $R_i$ and $R_f$ increases.  For example, the maximum penetration depth of the red absorption into the continuum, $D_{\rm max}$,  increases from 10\% to 30\% for $R_0=2R_*$ between an undiluted and a $f'=0.05$ flow. As before,  larger $R_0$ also increases the areal coverage and thus the depth and breadth of the absorption: For $R_0=4R_*$ and $F=0.2$, $D_{\rm max}$ reaches to 50\% of the stellar continuum for all viewing angles. In the unrealistic case of $R_0=8R_*$ and $F=0.1$, where $R_f$ extends to $35~R_*$, $D_{\rm max}$ can be 70\% of the stellar continuum.  However, for flows confined to maximum sizes on the order of corotation, the deepest penetrations are about 50\% of the stellar continuum. \\

\subsubsection{Further Comparison to Observations}

\begin{figure}
\epsscale{1.2}
\plotone{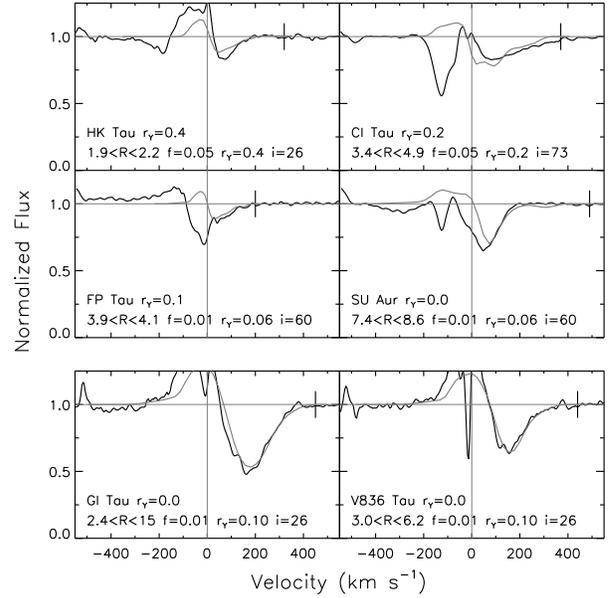} 
\figcaption{Examples of least-squares fits of dipolar model profiles (gray) to selected observations (black), with the stellar escape velocities marked by short vertical lines.   The top four panels use dipoles with $F=f$, and the corresponding $f$, $r_Y$, and $i$ are shown.  The bottom two panels have strong red absorptions and no detected veiling; they are well fit by extended dilute dipolar models ($R_0=4R_*$, $F=0.2$ for GI Tau; $R_0=4R_*$, $F=0.1$ for V836 Tau). Since processes other than scattering by the accretion flow can be important at low velocities, points with $V/V_{\rm esc}\le0.1$ are ignored in the fitting procedure. \label{f.fits}}
\epsscale{1}
\end{figure}

By introducing the concept of a diluted dipole, where field lines carrying accreting gas only dilutely fill the volume occupied by a wide magnetosphere, we can simultaneously generate deep and broad red absorption features while maintaining small filling factors for hot accretion shocks with $f\sim0.01$.  This is the empirical regime in Figure~\ref{f.comp} where dipolar models with $f=F$ were unable to account for stars with both strong absorption (large $W_\lambda'$) and very low veiling ($r_Y\sim0$).  We compare observed and model profiles for a few individual stars in Figure~\ref{f.fits}, in 4 cases for undiluted, fairly narrow dipoles for stars with $r_Y$ ranging from 0 to 0.4 and in 2 cases for wide, diluted dipoles with $r_Y\sim0$, where we have rescaled the model profiles to the escape velocity of each star.   Since we have not computed a large grid of models,  the magnetospheric properties listed for each fit are not intended to be predictions for a particular star.  However, this fitting procedure shows that weaker red absorptions can be reasonably described by basic undiluted models with a small range of origination radii in the disk, where veilings $r_Y$ from 0 to 0.4 can be consistently modeled with an appropriate choice of $f$, and the red absorption can include scattering contributions from both stellar and accretion shock continua.  Similarly, strong red absorptions in stars with low veilings can be well fit by dilutely filled flows with small $f$ but a wide span of origination radii in the disk, resulting in a large projected area of accreting gas for the scattering of the stellar continuum.  

\begin{figure}
\epsscale{1.2}
\plotone{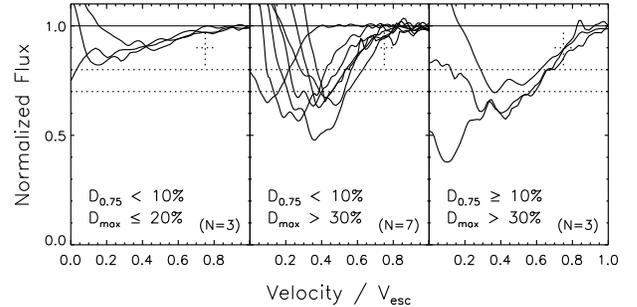}
\figcaption{Superposed \helium\ lines from the reference sample for the 13 stars with $r_Y\le0.1$, appropriate for modeling with dilute dipole flows with $f=0.01$. Only the red half of the profile is shown, normalized to the individual escape velocity of each star.  Profiles are grouped by $D_{0.75}$, the penetration depth into the continuum at $V/V_{\rm esc}=0.75$, and by $D_{\rm max}$, the maximum penetration into the continuum.  For comparison with models, the dotted cross in each panel marks $D_{0.75}=10\%$, and the dotted horizontal lines mark depths of 20\% and 30\%. \label{f.normed}}
\epsscale{1}
\end{figure}

The overall applicability of the diluted dipolar model can be appreciated by comparing the model profiles from Figure~\ref{f.dilute} to the ensemble of observed helium profiles for those stars with $r_Y\le0.1$ and thus $f \sim 0.01$, where the effect of scattering from a hot accretion shock will be inconsequential and  the properties of the red absorption will be shaped almost entirely by scattering of the stellar continuum. To effect this comparison in a general way, rather than focusing on individual stars, in Figure~\ref{f.normed} we plot superposed observed profiles for the redward side of \helium, each normalized to their respective escape velocity and separated into 3 groups on the basis of their depths both at 0.75~\vesc\ ($D_{0.75}$) and at maximum absorption ($D_{\rm max}$).  To aid in the comparison, both  Figures~\ref{f.dilute} and \ref{f.normed} denote depths for $D_{0.75}=10\%$ and $D_{\rm max}= 20\%$ and 30\%. 

In Figure~\ref{f.normed} the left panel contains the 3 shallowest profiles, with $D_{\rm max}\le20$\% and
$D_{0.75}<10$\%.  Compared to the predicted profiles in Figure~\ref{f.dilute}, the model flows that most resemble such broad but shallow profiles have $R_0\sim2R_*$ and $F\lesssim0.1$, although some viewing angles for larger flows with relatively small areal coverage of magnetic footpoints, $F\lesssim0.05$, could also apply.  Our coverage of parameter space is not exhaustive, but it is clear that for the broad but shallow red absorptions the range of radii over which the accretion flow leaves the disk is narrow, corresponding to a fairly small area on the star for the magnetospheric footpoints, but still larger than 1\%.  The central and right panels contain the 10 deeper profiles among the stars with low
veiling, where $D_{\rm max}$ ranges from 30 to 60\%.   Model flows that produce deeper profiles generally have significant areal coverage of magnetic footpoints $F$, as seen in Figure~\ref{f.dilute} where the accretion flow leaves the disk over a wide range of radii, impacting the star over a wide range of angles, in some cases with magnetic footpoint coverage up to 20\% of the stellar surface area. This is considerably larger than has been modeled in previous work on magnetospheric infall.

Even with wide diluted flows, the profiles of the 3 stars in the right panel of Figure~\ref{f.normed} (AA Tau, DK Tau, DN Tau) are a challenge to explain under the constraints of a dipolar geometry.  These profiles not only have $D_{\rm max}>30$\% but also have $D_{0.75}\ge10$\%, with the caveat that errors in escape velocity may be up to 20\%.   From the models explored in Figure~\ref{f.dilute}, flows with very wide extents, leaving the disk over a range of radii from a few $R_*$ to beyond corotation and viewed fairly close to pole-on, are required to produce profiles with $D_{0.75}\ge10$\%.  Rather than postulate an enormous dipolar flow with a polar viewing angle (which is clearly not the case for, at least, the edge-on source AA Tau), in the next section we will explore an example of a non-dipolar geometry to find a more plausible explanation for these three observations.

Only 8 CTTS in the reference sample have $r_Y>0.1$, such that the properties of the \helium\ red absorption may be affected by scattering of continuum photons from the hot accretion shock. One of these is DR Tau, where the high $r_Y=2$ implies $f\approx0.24$ (eq.~[\ref{e.veil}]), which as shown in Section 4.1 would yield a red absorption at least an order of magnitude stronger than the observed $W_\lambda'=0.05$. As will be addressed in Section 5, we suspect that in this case the red absorption has been filled in by a wind exterior to the accretion flow. 

\subsection{Diluted Radial Flows}

We have identified the 3 stars in the right panel of Figure~\ref{f.normed}, AA Tau, DK Tau, and DN Tau,  as difficult to explain with scattering in a dipolar geometry due to their absorption depths at velocities in excess of 0.5 \vesc.  In a dipolar flow, the impact velocity at the stellar surface depends on the polar angle $\theta$, which is determined by the initial distance of infall $R$ (eq.~[\ref{e.dipole}]), such that the impact velocity is greatest when $\theta$ is near the pole (i.e., $R$ is large) and diminishes as $\theta$ approaches the equator (i.e., $R$ becomes small).  Thus if $\theta$ is small enough, high impact velocities will result, although flows with small $\theta$ become highly curved and pinched as they reach the star (Fig.~\ref{f.geom}), resulting in small areal coverage and thus a shallow absorption profile at the highest velocities.

\begin{figure*}
\includegraphics[angle=90,width=\textwidth]{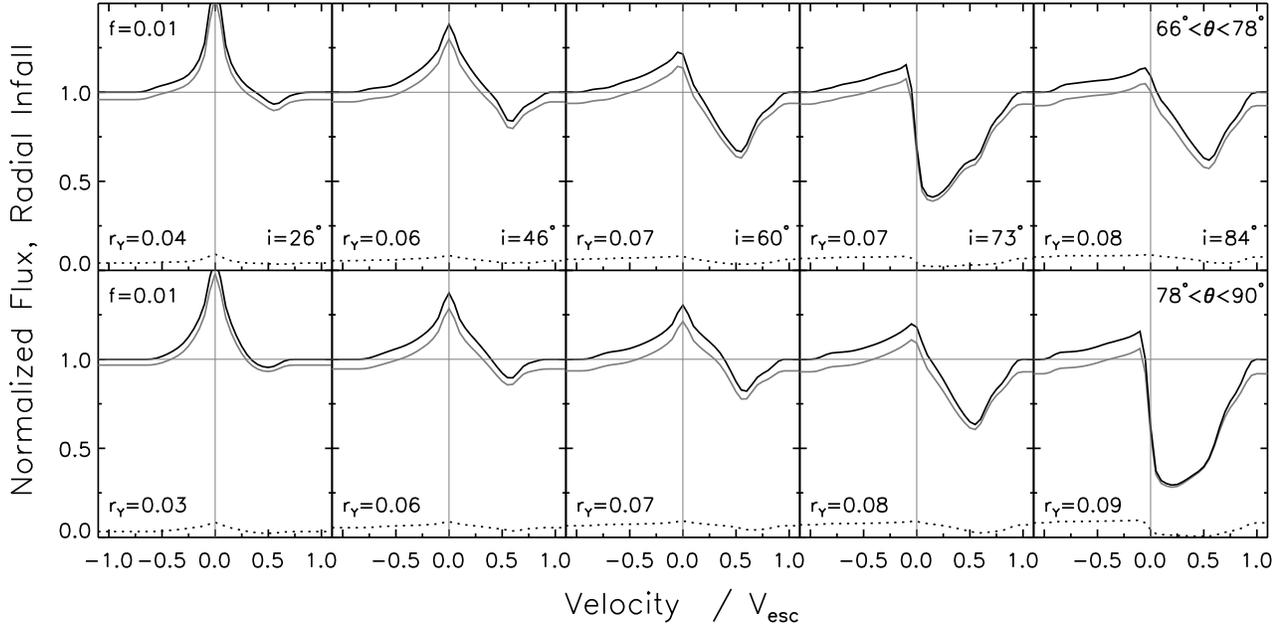}
\vskip -1.42in
\figcaption{Scattering profiles for diluted radial infall in non-rotating, azimuthally symmetric flows that begin at 8~$R_*$ and impact the star over a range of polar angles $\theta$ that encompasses $F=20\%$ of the stellar surface area.  Accreting field lines and their accretion shocks fill only 5\% of $F$ (i.e., $f=1\%$), with 1-$\micron$ veilings $r_Y$ as listed.  In the top row, the flow impacts the star over the range $66^\circ<\theta<78^\circ$, while in the bottom row, the flow impacts the star over the range $78^\circ<\theta<90^\circ$.  The same five viewing angles are used as in previous figures.\label{f.radial}}
\end{figure*}

We investigate radial infall trajectories as an alternative geometry that could produce deep absorption at high velocities.  In all aspects except the geometry,  radial models have the same assumptions as our dipolar models except that we have not included rotation.  The axisymmetric flow begins at some distance from the star $R_{\rm max}$, and it falls radially toward the star, impacting the stellar surface between polar angles $\theta_1$ and $\theta_2$ in one hemisphere and between $\pi-\theta_1$ and $\pi-\theta_2$ in the other.  The fractional surface of the star spanned by the accretion flow, $F$, is $\cos\theta_1-\cos\theta_2$, and the shocks within this region together occupy a fraction $f'$ of the area $F$, so that $f=Ff'$.  The disk truncation radius is a free parameter, but we set it equal to $R_{\rm max}$, which is 8~$R_*$ in all radial models. Figure~\ref{f.radial} shows scattering profiles from two radial geometries at 5 viewing angles.  In the top row, the impact region extends from $\theta_1=66.4^\circ$ to $\theta_2=78.5^\circ$, while in the bottom row, the impact region extends from $\theta_1=78.5^\circ$ to the equator.  In both cases, $F=0.2$ and $f=0.01$.  As expected, the absorption is strongest for a viewing angle within the confines of the flow (i.e., $\theta_1<i<\theta_2$), and the profile becomes a nearly symmetric emission profile (assuming axisymmetry and no rotation) for views close to pole-on. When the viewing angle is aligned or nearly aligned with the column of absorbing gas, each radial model can produce the observed range of absorption depths at high velocities, with $D_{0.75}>10$\% for profiles with $i>60^\circ$ in the top row and $i>73^\circ$ in the bottom row.

We are not advocating radial infall starting from a large distance, and thus, the profile sequences in Figure~\ref{f.radial} are not expected to be realistic for the whole velocity range. However, the requisite deep absorption at high velocities, resulting from material moving faster than $\sim2/3~V_{\rm esc}$, all arises inside of about 2~$R_*$. Thus the message from these calculations is that the accretion stream only need move in a radial trajectory, i.e., become less curved than a dipole, as it nears the star.  

\begin{figure}
\epsscale{1.2}
\plotone{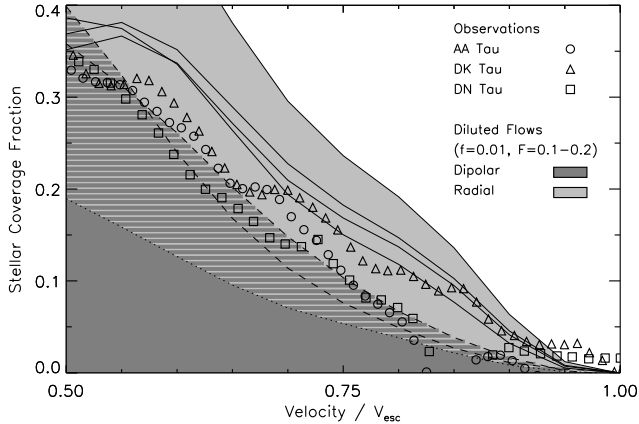}
\figcaption{High-velocity tails of observed and model profiles for 3 stars with the largest values of $D_{0.75}$ and $r_Y=0$, inverted to show the minimum fraction of the star occulted by infalling material at each velocity.  Dark shading indicates the regime of diluted dipolar models with $f=0.01$ and infall contained entirely within a typical corotation radius, marked by the profile (dotted line) with $F=0.1$, $3.0<R/R_*<6.2$, and $i=46^\circ$.   The hatched region indicates the extension when dipolar field lines out to $\sim$ twice the corotation radius participate in infall, marked by the profiles (dashed lines) with $F=0.2$, $2.4<R/R_*<15$, and $i=26^\circ$ or $46^\circ$.  Light shading shows the regime for profiles formed in diluted radial infall with $F=0.2$ and $f=0.01$.  For both radial geometries in Figure~\ref{f.radial}, two profiles (solid lines) with $i$ close to the infall angle are shown.\label{f.oplot}}
\epsscale{1}
\end{figure}

The effectiveness of radial infall trajectories for material near the star in accounting for the high-velocity absorption in AA Tau, DK Tau, and DN Tau is shown in Figure~\ref{f.oplot}. The figure shows model and observed profiles where (1) profiles are inverted so the vertical axis is a measure of the minimum stellar coverage fraction at each velocity and (2) only velocities in excess of 0.5~\vesc\ are plotted. The regime of diluted dipolar models with the largest $D_{0.75}$ is shown with dark and hatched shading, while the regime of flows with radial trajectories for gas near the star is shown with light shading.  The dark shading is for the best case from our diluted dipolar models for a flow contained entirely within the corotation radius: $F=0.1$ originating  between 3.0 and 6.2~$R_*$ in the disk and viewed from $i=46^\circ$. Although 8~$R_*$ is a more typical corotation radius, extending the flow out to this distance would not produce much additional absorption. The hatched region is for a diluted dipole that allows field lines extending out to nearly twice the corotation radius to participate in the flow, where the dashed lines are for the case $R_0=4R_*$, $F=0.20$ ($2.4<R/R_*<15$), seen from two viewing angles, $i=26^\circ$ and $i=46^\circ$.  Although the latter two extreme dipolar models come close to producing sufficient absorption at high velocities, significant accretion beyond corotation is likely not physical. In contrast to the dipole trajectories, the regime of the four radial models from Figure~\ref{f.radial} with viewing angles nearly aligned with the infalling gas easily contains the observed stellar coverage fraction from 0.6 to 0.85~\vesc, with no requirement that the flow originate at radii beyond corotation. The realistic situation is likely to involve some complex magnetic field topologies with trajectories approaching radial as they near the star.

\section{DISCUSSION\label{s.discuss}}

\subsection{Implications of Diluted Funnel Flows}

The high opacity and resonance scattering properties of \helium\ enable the geometry of magnetospheric accretion to be probed via absorption of gas seen in projection against the star, in contrast to previous studies that rely on the morphology of emission lines. Under the assumptions that the flow is an azimuthally symmetric dipole and helium is sufficiently optically thick that all incident 1-$\micron$ radiation is scattered, we have illustrated the sensitivity of the red absorption to both the angular extent of the magnetosphere and the filling factor of hot gas from the accretion shock $f$. If $f$ exceeds a few percent, the hot spot will be an important contributor to the scattering of the 1-$\micron$ continuum; however, since the strongest and broadest \helium\ absorptions are seen in stars with little or no 1-$\micron$ veiling, these red absorptions must instead arise almost solely by scattering of photospheric radiation. Achieving the observed breadth and depth of the absorption requires a large angular coverage of the stellar continuum in the azimuthal direction over a wide range of velocities for many stars, with areal coverage in footpoints on the star of  $F=10-20\%$. We suggest that the required combination of wide flows and low filling factors of hot gas is a result of accretion in many narrow streamlets, each of which may have a dipolar configuration but which together only fill a small fraction of the enclosed volume. We have explored the case where the streamlets are uniformly distributed through the accreting volume, producing  wide, dilutely filled flows that reconcile the need for absorption over a broad range of velocities with filling factors of hot gas $f<1\%$, as observed (CG). 

Earlier studies also imply a discrepancy between the areal coverage $F$ of magnetospheric footpoints and the filling factor of hot accretion shocks $f$. For example, magnetospheres with $f=F=8\%$ were invoked to model hydrogen lines arising from accretion flows in order to produce sufficient line fluxes and mass accretion rates \citep{sym05,kur06}.   The seminal sequence of papers modeling hydrogen line formation in funnel flows from \citet{har94} to \citet{muz01} also required filling factors that were larger than predicted by SED modeling of continuum excesses to account for observed emission line luminosities.  The notion of accretion via streamlets that dilutely fill a large volume is a straightforward way to reconcile this discrepancy, simultaneously allowing large field sizes and small shock filling factors.  

Although our model invokes diluted accretion flows in widely distributed streamlets of gas, an alternate scenario for diluted accretion is the one suggested by the MHD simulations of \citet{rom04}, where internal structure within the accretion flow gives a mass accretion rate (and a corresponding blackbody continuum temperature) that is highest at the interior and falls off toward the sides.  Although this scenario can also, for a large $F$, produce a smaller veiling from the area-weighted blackbody continua than the undiluted $f=F$ case, the advantage to widely dispersed streamlets is that they provide a facile means for ionizing radiation to penetrate to most of the infalling gas, since distributed accretion shocks with small individual areas would allow ionizing photons produced in each shock to escape more easily from the sides and ionize helium at other locations.  Another consequence of such distributed accretion shocks is that photons from the shocks emitted toward the star would be incident on a larger area of the photosphere than for a single shock with the same $f$. This may invalidate the usual assumption of a plane-parallel geometry for the radiative transfer of photons with the effect that, independent of the internal structure within an individual streamlet, the resultant veiling continuum would encompass a range in blackbody temperatures.  In a wide flow where dilution is somewhat uniform, there will be many separate shocks with a range of blackbody temperatures surrounding them.  There may be some observational support for this phenomenon in that the veiling continuum longward of 0.5~$\micron$ (\citealt{bas90}; \citealt{whi04}; EFHK) is broader than the single 8000-K blackbody that is a good match to the excess at shorter wavelengths (CG). 

Constraints on the angular extent of accreting gas and the location in the disk where infall originates are relevant to models for disk locking and wind launching. Although there are some cases where \helium\ profiles resemble those expected from viewing an accretion funnel restricted to a narrow origination around the corotation radius, the suite of profiles expected from viewing this magnetic topology from all inclination angles is not consistent with the observations.  The most extreme deep and broad absorptions instead require infall spanning a wide extent of origination radii, from a few $R_*$ out to at least typical corotation radii of 6 to 8~$R_*$ if the flows are dipolar.  For other magnetic field configurations, such as a tilted dipole or a multipole field, significant red absorption need not require such a wide range in initial infall distances.  In general, the depth of the red absorption is governed more by the range of impact latitudes than by the range of initial radii; only for a dipolar flow aligned with the rotational axis are the two ranges so closely linked.  For example, in an aligned dipolar flow with $R_0=4R_*$ and $F=0.1$, the range in impact latitude of 23.6-35.3$^\circ$ corresponds to a range in initial radius of 3.0-6.2~$R_*$.  In a more complex magnetic configuration, a comparably wide range of impact angles could produce a strong red absorption without the need for such a large range of initial radii.

The necessity for a dilutely filled flow does imply that there is not a sharp delineation on the disk for accretion onto the star. It likely indicates a very inhomogeneous field structure at large distances, with many local pockets distributed over a broad radial range on the disk giving rise to accretion streamlets.  Since our analysis assumes axial symmetry in a set of nested dipolar flows, the constraints that the breadth and depth of the red absorption place on the angular extent of the accreting gas are even more extreme if, as likely, accretion channels are in restricted azimuth zones.   Furthermore, there are some red absorptions that are so deep at velocities $\ge 0.5~V_{\rm esc}$ that a dipole morphology is inadequate, even when arising from 2 $R_*$ to the corotation radius. In these cases we find that radially directed infall can achieve the requisite depth of absorption, although other topologies that result in a large covering factor of the star at the highest velocities can likely be constructed. 

Recent Doppler tomographic maps of the CTTS V2129 Oph and BP Tau, based on circular polarization of \ion{Ca}{2} \citep{don07,don08}, reveal the locations of accretion hot spots on the stars.  The spots span quite a broad latitude range (extending roughly from the pole to $45^\circ$) but a very narrow azimuthal range.  The narrow azimuthal range implies that the detection of \helium\ red absorption requires an opportune time at which the accretion spots are directly in view.  This situation is consistent with the result that BP Tau, a mildly accreting CTTS included in our \helium\ survey, did not show any \helium\ red absorption on the two occasions we observed it.  At present, there are not enough tomographic data to see how consistent this pattern is among a range of accreting stars, although our detection of subcontinuum red absorption in 21 of 38 CTTS, including 20 of 29 stars (and 37 of 56 total spectra) with $r_Y\le0.5$ ($r_V\le2$), would imply a large azimuthal coverage by the accretion spots.  However, we note that even in the two stars with tomographic maps, there is a possibility that accretion impacts the stars over a wide range of longitudes.  Donati et al.\ attribute only 2/3 of the \heopt\ emission but all of the \heopt\ circular polarization to accretion spots, based on the fraction of the emission that shows rotational modulation compared to that which is time-independent.  The time-independent component, responsible for 1/3 of the \heopt\ emission, is attributed to a chromospheric component distributed uniformly over the stellar surface.  However, since non-accreting WTTS show either very weak or, more commonly, no \heopt\ emission \citep{ber01}, it would appear that TTS chromospheres are not significant contributors to this line.  Instead, the time-independent component may be from more widely distributed accretion shocks that cover a broader range of longitudes.

\subsection{Absence of \helium\ Red Absorption}

In this paper we have focused on the 21/38 CTTS that show redshifted absorption in \helium\ at least once in an observational program with sporadic time coverage. Clearly the absence of \helium\ red absorption is also important in constraining the topology of magnetospheric accretion. An important point is that \helium\ red absorption is rarely seen among CTTS with the highest 1-$\micron$ veiling (1/25 observations; see Section 3.1).  Of the 9 stars in the EFHK survey in this category, the only one that showed redshifted absorption, on 1 of 4 occasions, is DR Tau. We suspect that in all 9 of these stars, emission from a wind exterior to the accretion flow, instead of from the flow itself, is filling in any redshifted absorption that may be present.  If in-situ emission from the funnel flow were significant, it would be difficult for it to fill up the absorption at the red edge of the profile, since the geometry of the funnel flow results in smaller volumes at higher velocities, producing centrally peaked emission profiles that fall off rapidly toward both blue and red high velocities (see the contribution to the emission from scattering of the stellar continuum in Fig.~\ref{f.example}).  The near absence of red absorption among these stars instead calls for a situation in which the redshifted absorption, if present, is filled in completely. In the case of DR Tau, it is clear that weak red absorption, confined to high velocities, is visible when the emission from the P Cygni wind profile is weakest (see Fig.~\ref{f.multi}). Among the other stars in this high-veiling group, all have either broad blue helium absorptions indicative of viewing through a stellar wind or strong helium emission interpreted to arise in a conical stellar wind viewed obliquely (see KEF and EFHK). Either of these contributions to redward emission would be sufficient to fill up even a strong red absorption, provided the wind is optically thick and exterior to the accretion flow.

When profiles from both \helium\ and \heopt\ are considered, the evidence suggests that \helium\ red absorption is rare or absent in CTTS with large 1-$\micron$ veiling not primarily because the absorption is being filled in by wind emission but more because the geometry of the funnel flows is altered compared to that of low-veiling CTTS.  This inference is drawn from a study of \heopt\ profiles and optical veiling presented in \citet{ber01}, which includes many stars in common with EFHK.  They found that CTTS whose \heopt\ profiles showed only a narrow component, consistent with formation in post-shock gas from an accretion shock, show an excellent correlation between the strength of narrow-component helium emission and optical veiling. In contrast, CTTS whose \heopt\ profiles show a contribution from a broad component have reduced or absent emission from a narrow component relative to stars of similar optical veiling.  While it might appear otherwise, this is not an esoteric point!  It suggests that CTTS with strong stellar winds and high optical and 1-$\micron$ veilings may have crunched or otherwise-altered magnetospheres resulting in weak narrow-component emission from a hot accretion shock, and there is a significant contribution to the veiling continuum from another source. In contrast, CTTS without strong stellar winds (many of which show disk wind profiles at \helium; see KEF) have extensive magnetospheres carrying accreting gas to the star, and hot accretion shocks are the dominant contributor to their optical veiling. We anticipate being able to test this suggestion shortly, following analysis of simultaneously obtained spectra extending from 0.4 to 2.2~$\micron$. 

A second point regarding the frequency of \helium\ red absorption is that, in contrast to CTTS with high 1-$\micron$ veiling, red absorption is commonly seen in stars with lower veiling (37 out of 56 spectra for $r_Y \le 0.5$; see Section 3.1).  Among this group, some objects (e.g., TW Hya and CY Tau; see Fig.~\ref{f.multi}) clearly show reduction of the red absorption as the emission, likely that of a stellar wind as indicated by the strong P Cygni profile, increases. In such stars the appearance and disappearance of the red absorption is likely due, at least in part, to filling in by an exterior stellar wind, as in DR Tau. In others (e.g., V836 Tau and GK Tau; Fig.~\ref{f.multi}), the weaker helium emission could arise simply from scattering in the funnel flow, and the absence of red absorption may indicate viewing at an azimuth with no funnel flow activity. Azimuthal asymmetry in the funnel flow is also the likely explanation for the strongly variable red absorption morphology in objects such as AA Tau and DK Tau (Fig.~\ref{f.multi}). 

The possibility that red absorption may be partially filled in, either by in-situ emission from the accretion flow or by scattered or in-situ emission from a wind, implies that the true magnitudes of some red absorptions are stronger and their constraints on the flow structure stiffer than the observations indicate.  Further, since red absorption can be completely filled in by in-situ wind emission or, in some cases, not be observed at all due to azimuthal inhomogeneities, \helium\ red absorption is likely more pervasive among CTTS than is already apparent.

\subsection{Size and Structure of the Accretion Flow}

Inferences to date on the physical extent of accretion flows have largely relied on models positing that hydrogen and sodium lines are formed primarily in these flows \citep{cal00}. A correlation between the emitting area of the accretion flow and the magnitude of the mass accretion rate has been suggested by \citet{muz01} as the explanation for the well-established empirical correlation between infrared hydrogen line luminosities and accretion luminosities \citep{muz98,fol01,nat04}.  The models of hydrogen line formation in magnetospheric flows  predict that hydrogen line luminosities are primarily determined by the surface area of the accreting gas, not the density in the flow. The suggestion is that objects with higher accretion rates require larger emitting areas for their magnetospheres than objects with smaller accretion rates. Since more extended magnetospheres are expected on theoretical grounds in objects with lower disk accretion rates, a further suggestion is that high-accretion-rate objects have wider azimuthal coverage of accreting columns.  The red absorption profiles of \helium\ give new insight into this phenomenon, since we have a clear indication of very extended and wide flows in stars with low accretion rates. For example, our limited phase coverage of the edge-on system AA Tau shows that at the same time exceptionally strong red absorption at \helium\ is observed, requiring extensive but dilutely filled accretion flows, the hydrogen P$\gamma$ profile is weak, narrow, and symmetric, suggesting a small magnetospheric emitting area if it is formed in the accretion flow.  We anticipate that time-monitoring campaigns combining profile monitoring of both \helium\ and the immediately adjacent P$\gamma$ line of hydrogen will provide a definitive assessment of the size and azimuthal coverage of the funnel flow and possibly also clarify the origin of the correlation between infrared hydrogen line luminosities and the accretion luminosity.

\section{CONCLUSIONS}

We have probed the geometry of magnetospheric accretion in classical T Tauri stars by modeling red absorption at \helium\ via scattering of the stellar and veiling continua.  Between 2001 and 2007, we acquired 81 1-$\micron$ spectra of 38 CTTS spanning the full observed range of mass accretion rates.  Of the 38 stars, 1 of 9 with $r_Y>0.5$ and 20 of 29 with $r_Y\le0.5$ show red absorption at \helium\ that extends below the 1-$\micron$ continuum in one or more spectra, demonstrating that red absorption from magnetospheric accretion is rare in objects with high veiling but is found in about two thirds of objects with moderate to low veiling. The red absorptions can be strong, deep, and broad, with equivalent widths up to 4.5~\AA, maximum penetrations into the 1-$\micron$ continuum up to 61\%, and widths at one quarter of the absorption minimum up to 320~\kms; furthermore, they tend to be strongest in stars with the lowest veilings.

We model the red absorption by assuming that an axisymmetric dipolar accretion flow scatters photons from the star and from hot zones in the accretion-heated photosphere that produce the 1-$\micron$ veiling and have filling factor $f$.  Testing a range of magnetosphere widths and $f$ consistent with shock filling factors from the literature, we find that about half of the absorption profiles can be explained by dipolar flows in which the size of the flow is consistent with the size of the shock filling factor $f$. Weak absorptions in stars with weak veiling and intermediate absorptions in stars with intermediate veiling are explained by such flows, but strong absorptions in stars with little to no veiling are not.

We introduce the concept of dilution as a means of producing a strong red absorption while keeping the filling factor and thus the veiling low.  In a diluted flow, the magnetosphere can extend over a wide range of radii, with a large covering factor on the stellar surface, but this volume is incompletely filled by accreting gas.  Instead of a single thick flow, we posit  multiple nested streamlets with a total filling factor  small enough for a low veiling, but each with an intrinsic thermal or turbulent width sufficient to scatter photons as though the entire volume were filled, thereby yielding a large red absorption.  The multiple streamlets can also explain how helium is ionized through the entire flow, rather than just the skin of a thick flow.  Large, dilutely filled accretion flows are necessary for about half of the objects, some of which require accreting streamlets to connect to the disk over a range from 2~$R_*$ out to or beyond corotation. A few stars show such deep absorption at redward velocities exceeding 50\% of the stellar escape velocity that flows near the star with less curvature than a dipolar trajectory seem to be required.  

The frequency of \helium\ red absorption is also informative. Our limited temporal coverage suggests that the frequency of helium absorption differs in stars with high and low veiling.  Red absorption at \helium\ is far more common in stars with low veiling.  When it is absent from these stars, it is sometimes because helium emission from another source such as a wind fills it in and sometimes because of  inhomogeneous azimuthal coverage of accreting magnetic columns.  Among stars with high veiling ($r_Y\ge0.5$), red absorption at \helium\ is rarely seen. If these stars had accretion geometries similar to those of the low-veiling stars, they would be expected to have extremely strong red absorptions.  Even if the absorption were filled in by emission from the accretion flow, the stars would still be expected to show red absorption at high velocities. In the high-veiling stars, the paucity of \helium\ red absorption, the presence of \helium\ emission and blue absorption that suggest formation in accretion-powered stellar winds, and the weakness or absence of narrow-component \heopt\ emission from an accretion shock lead us to suggest that the magnetospheric accretion structure may be crunched or otherwise reduced in CTTS with the highest disk accretion rates.

We find the study of \helium\ red absorption due to infalling gas projected in front of the star to be complementary to studies of emission lines modeled as arising over the full size of the accretion flow. The proximity of \helium\ and P$\gamma$ offer an excellent pair of lines for deeper investigation of magnetospheric geometries through intensive time-monitoring programs that can track non-aziumuthal structures as stars rotate. Our limited phase coverage of AA Tau demonstrates that this approach will be very effective, particularly when coupled with radiative transfer models that can constrain formation conditions for both lines simultaneously.

\acknowledgments

NASA grant NNG506GE47G issued through the Office of Space Science provides support for this project.  Thanks to A. Rostopchina for personally providing the last measurement needed to derive stellar parameters for every star in the sample and to M. Romanova for stimulating conversations on accretion flows.  We acknowledge helpful conversations with J. Bjorkman, S. Cabrit, N. Calvet, L. Hartmann, S. Matt, and an anonymous referee. The authors wish to recognize and acknowledge the very significant cultural role and reverence that the summit of Mauna Kea has always had within the indigenous Hawaiian community.  We are most fortunate to have had the opportunity to conduct observations with the Keck II telescope from this mountain.


\begin{thebibliography}{}

\bibitem[Alencar \& Basri(2000)]{ale00}
Alencar, S. H. P., \& Basri, G.  2000, \aj, 119, 1881

\bibitem[Basri \& Batalha(1990)]{bas90}
Basri, G., \& Batalha, C.  1990, \apj, 363, 654

\bibitem[Beristain et al.(2001)]{ber01}
Beristain, G., Edwards, S., \& Kwan, J.  2001, \apj, 551, 1037

\bibitem[Bertout et al.(1996)]{ber96}
Bertout, C., Harder, S., Malbet, F., Mennessier, C., \& Regev, O.  1996, \aj, 112, 2159 

\bibitem[Bouvier et al.(2007a)]{bou07a}
Bouvier, J., Alencar, S. H. P., Harries, T. J., Johns-Krull, C. M., \& Romanova, M. M.  2007a, in Protostars and Planets V, ed. B. Reipurth, D. Jewitt, \& K. Keil (Tucson: Univ. Arizona Press), 479

\bibitem[Bouvier et al.(1986)]{bou86}
Bouvier, J., Bertout, C., \& Bouchet, P.  1986, \aap, 158, 149 

\bibitem[Bouvier et al.(1993)]{bou93}
Bouvier, J., Cabrit, S., Fernandez, M., Martin, E. L., \& Matthews, J. M.  1993, \aap, 272, 176 

\bibitem[Bouvier et al.(1995)]{bou95}
Bouvier, J., Covino, E., Kovo, O., Martin, E. L., Matthews, J. M., Terranegra, L., \& Beck, S. C.  1995, \aap, 299, 89 

\bibitem[Bouvier et al.(1999)]{bou99}
Bouvier, J., et al.  1999, \aap, 349, 619

\bibitem[Bouvier et al.(2003)]{bou03}
--------.  2003, \aap, 409, 169

\bibitem[Bouvier et al.(2007b)]{bou07b}
--------.  2007b, \aap, 463, 1017

\bibitem[Calvet \& Gullbring(1998)]{cal98}
Calvet, N., \& Gullbring, E.  1998, \apj, 509, 802 (CG)

\bibitem[Calvet \& Hartmann(1992)]{cal92}
Calvet, N., \& Hartmann, L.  1992, \apj, 386, 239

\bibitem[Calvet et al.(2000)]{cal00}
Calvet, N., Hartmann, L., \& Strom, S. E.  2000, in Protostars and Planets IV, ed. V. Mannings, A. P. Boss, \& S. S. Russell (Tucson: Univ. Arizona Press), 377

\bibitem[Collier Cameron \& Campbell(1993)]{col93}
Collier Cameron, A., \& Campbell, C. G.  1993, \aap, 274, 309

\bibitem[Correia et al.(2006)]{cor06}
Correia, S., Zinnecker, H., Ratzka, Th., \& Sterzik, M. F.  2006, \aap, 459, 909

\bibitem[DeWarf et al.(2003)]{dew03}
DeWarf, L. E., Sepinsky, J. F., Guinan, E. F., Ribas, I., \& Nadalin, I.  2003, \apj, 590, 357 

\bibitem[Donati et al.(2007)]{don07}
Donati, J.-F., et al.  2007, \mnras, 380, 1297

\bibitem[Donati et al.(2008)]{don08}
--------.  2008, \mnras, 386, 1234

\bibitem[Edwards et al.(2006)]{EFHK}
Edwards, S., Fischer, W., Hillenbrand, L., \& Kwan, J.  2006, \apj, 646, 319 (EFHK)

\bibitem[Edwards et al.(1994)]{edw94}
Edwards, S., Hartigan, P., Ghandour, L., \& Andrulis, C.  1994, \aj, 108, 1056

\bibitem[Ferreira et al.(2006)]{fer06} 
Ferreira, J., Dougados, C., \& Cabrit, S.  2006, \aap, 453, 785

\bibitem[Folha \& Emerson(2001)]{fol01}
Folha, D. F. M., \& Emerson, J. P.  2001, \aap, 365, 90

\bibitem[Ghez et al.(1993)]{ghe93} 
Ghez, A. M., Neugebauer, G., \& Matthews, K.  1993, \aj, 106, 2005

\bibitem[Ghosh \& Lamb(1978)]{gho78}
Ghosh, P., \& Lamb, F. K.  1978, \apj, 223, L83

\bibitem[Gregory et al.(2006)]{gre06} 
Gregory, S. G., Jardine, M., Simpson, I., \& Donati, J.-F.  2006, \mnras, 371, 999

\bibitem[Guenther \& Hessman(1993)]{gue93}
Guenther, E., \& Hessman, F. V.  1993, \aap, 276, L25

\bibitem[Gullbring et al.(2000)]{gul00}
Gullbring, E., Calvet, N., Muzerolle, J., \& Hartmann, L.  2000, \apj, 544, 927

\bibitem[Gullbring et al.(1998)]{gul98}
Gullbring, E., Hartmann, L., Brice\~{n}o, C., \& Calvet, N.  1998, \apj, 492, 323

\bibitem[Hartigan et al.(1995)]{har95}
Hartigan, P., Edwards, S., \& Ghandour, L.  1995, \apj, 452, 736 (HEG)

\bibitem[Hartigan et al.(1989)]{har89}
Hartigan, P., Hartmann, L., Kenyon, S., Hewett, R., \& Stauffer, J.  1989, \apjs, 70, 899

\bibitem[Hartigan \& Kenyon(2003)]{har03}
Hartigan, P., \& Kenyon, S. J.  2003, \apj, 583, 334

\bibitem[Hartigan et al.(1991)]{har91}
Hartigan, P., Kenyon, S. J., Hartmann, L., Strom, S. E., Edwards, S., Welty, A. D., \& Stauffer, J.  1991, \apj, 382, 617

\bibitem[Hartmann et al.(1994)]{har94}
Hartmann, L., Hewett, R., \& Calvet, N.  1994, \apj, 426, 669

\bibitem[Hillenbrand \& White(2004)]{hil04}
Hillenbrand, L. A., \& White, R. J.  2004, \apj, 604, 741

\bibitem[Johns-Krull(2007)]{joh07}
Johns-Krull, C. M.  2007, \apj, 664, 975

\bibitem[Johns-Krull et al.(2000)]{joh00}
Johns-Krull, C. M., Valenti, J. A., \& Linsky, J. L.  2000, \apj, 539, 815

\bibitem[Kenyon \& Hartmann(1995)]{ken95}
Kenyon, S. J., \& Hartmann, L.  1995, \apjs, 101, 117

\bibitem[K\"{o}nigl(1991)]{kon91}
K\"{o}nigl, A.  1991, \apj, 370, L39

\bibitem[Kurosawa et al.(2006)]{kur06}
Kurosawa, R., Harries, T. J., \& Symington, N. H.  2006, \mnras, 370, 580

\bibitem[Kurosawa et al.(2008)]{kur08}
Kurosawa, R., Romanova, M. M., \& Harries, T. J.  2008, \mnras, 385, 1931

\bibitem[Kwan et al.(2007)]{kwa07}
Kwan, J., Edwards, S., \& Fischer, W.  2007, \apj, 657, 897 (KEF)

\bibitem[Lawson \& Crause(2005)]{law05}
Lawson, W. A., \& Crause, L. A.  2005, \mnras, 357, 1399 

\bibitem[Long et al.(2007)]{lon07}
Long, M., Romanova, M. M., \& Lovelace, R. V. E.  2007, \mnras, 374, 436

\bibitem[Long et al.(2008)]{lon08}
--------.  2008, \mnras, 386, 1274

\bibitem[Matt \& Pudritz(2005)]{mat05}
Matt, S., \& Pudritz, R. E.  2005, \apj, 632, L135

\bibitem[Matt \& Pudritz(2008a)]{mat08a}
--------.  2008a, \apj, 678, 1109

\bibitem[Matt \& Pudritz(2008b)]{mat08b}
--------.  2008b, \apj, in press

\bibitem[McLean et al.(1998)]{mcl98}
McLean, I. S., et al.  1998, \procspie, 3354, 566

\bibitem[Menten et al.(2007)]{men07}
Menten, K. M., Reid, M. J., Forbrich, J., \& Brunthaler, A.  2007, \aap, 474, 515

\bibitem[Mohanty \& Shu(2008)]{moh08}
Mohanty, S., \& Shu, F.  2008, \apj, in press

\bibitem[Mora et al.(2001)]{mor01}
Mora, A., et al.  2001, \aap, 378, 116

\bibitem[Muzerolle et al.(2000)]{muz00}
Muzerolle, J., Calvet, N., Brice\~{n}o, C., Hartmann, L., \& Hillenbrand, L.  2000, \apj, 535, L47

\bibitem[Muzerolle et al.(2001)]{muz01}
Muzerolle, J., Calvet, N., \& Hartmann, L.  2001, \apj, 550, 944

\bibitem[Muzerolle et al.(1998)]{muz98}
Muzerolle, J., Hartmann, L., \& Calvet, N.  1998, \aj, 116, 2965

\bibitem[Natta et al.(2004)]{nat04}
Natta, A., Testi, L., Muzerolle, J., Randich, S., Comer\'{o}n, F., \& Persi, P.  2004, \aap, 424, 603

\bibitem[Rebull et al.(2004)]{reb04}
Rebull, L. M., Wolff, S. C., \& Strom, S. E.  2004, \aj, 127, 1029

\bibitem[Romanova et al.(2008)]{rom08}
Romanova, M. M., Kulkarni, A. K., \& Lovelace, R. V. E.  2008, \apj, 673, L171

\bibitem[Romanova et al.(2003)]{rom03}
Romanova, M. M., Ustyugova, G. V., Koldoba, A. V., \& Lovelace, R. V. E.  2003, \apj, 595, 1009

\bibitem[Romanova et al.(2004)]{rom04}
--------.  2004, \apj, 610, 920

\bibitem[Rostopchina(1999)]{ros99}
Rostopchina, A. N.  1999, Astronomy Reports, 43, 113 

\bibitem[Rydgren et al.(1984)]{ryd84}
Rydgren, A. E., Zak, D. S., Vrba, F. J., Chugainov, P. F., \& Zajtseva, G. V.  1984, \aj, 89, 1015 

\bibitem[Sauty \& Tsinganos(1994)]{sau94}
Sauty, C., \& Tsinganos, K.  1994, \aap, 287, 893

\bibitem[Shu et al.(1994)]{shu94}
Shu, F., Najita, J., Ostriker, E., Wilkin, F., Ruden, S., \& Lizano, S.  1994, \apj, 429, 781

\bibitem[Siess et al.(2000)]{sie00}
Siess, L., Dufour, E., \& Forestini, M.  2000, \aap, 358, 593

\bibitem[Symington et al.(2005)]{sym05}
Symington, N. H., Harries, T. J., \& Kurosawa, R.  2005, \mnras, 356, 1489

\bibitem[Valenti \& Johns-Krull(2004)]{val04}
Valenti, J. A., \& Johns-Krull, C. M.  2004, \apss, 292, 619

\bibitem[von Rekowski \& Brandenburg(2006)]{von06}
von Rekowski, B., \& Brandenburg, A.  2006, Astronomische Nachrichten, 327, 53 

\bibitem[Vrba et al.(1986)]{vrb86}
Vrba, F. J., Rydgren, A. E., Chugainov, P. F., Shakovskaia, N. I., \& Zak, D. S.  1986, \apj, 306, 199 

\bibitem[Walker(1972)]{wal72}
Walker, M. F.  1972, \apj, 175, 89

\bibitem[Webb et al.(1999)]{web99}
Webb, R. A., Zuckerman, B., Platais, I., Patience, J., White, R. J., Schwartz, M. J., \& McCarthy, C.  1999, \apj, 512, L63

\bibitem[White \& Ghez(2001)]{whi01}
White, R. J., \& Ghez, A. M.  2001, \apj, 556, 265

\bibitem[White \& Hillenbrand(2004)]{whi04}
White, R. J., \& Hillenbrand, L. A.  2004, \apj, 616, 998

\bibitem[Yang et al.(2007)]{yan07}
Yang, H., Johns-Krull, C. M., \& Valenti, J. A.  2007, \aj, 133, 73

\end{thebibliography}
\end{document}